\documentclass[sigconf,nonacm,pbalance]{acmart}

\usepackage{booktabs}
\usepackage{multirow}
\usepackage{tabularx}
\usepackage{flafter}
\usepackage{placeins}

\makeatletter
\AtBeginDocument{%
  \patchcmd{\@BAlancecol}{\unvbox1\vfil}{\unvbox1}{}{}%
  \patchcmd{\@BAlancecol}{\kern-\dimen2\vfil}{\kern-\dimen2}{}{}%
}
\makeatother

\setcopyright{none}
\acmConference{}{}{}
\acmDOI{}
\acmISBN{}

\begin{document}
\flushbottom

\title{DualEvolve: Dual-Level Self-Evolution of Mechanism and Orchestration for Recommendation}

\author{Tianqi Gao}
\authornote{Tianqi Gao and Chenxu Wang contributed equally to this work.}
\email{tianqig358@gmail.com}
\affiliation[obeypunctuation=true]{%
  \institution{Independent Researcher}
  \city{}
  \country{}
}

\author{Chenxu Wang}
\authornotemark[1]
\email{wangchenxu13@meituan.com}
\affiliation[obeypunctuation=true]{%
  \institution{Meituan}
  \city{}
  \country{}
}

\author{Wenxuan Yang}
\email{yangwenxuan03@meituan.com}
\affiliation[obeypunctuation=true]{%
  \institution{Meituan}
  \city{}
  \country{}
}

\author{Yueqi Mou}
\email{mouyueqi@meituan.com}
\affiliation[obeypunctuation=true]{%
  \institution{Meituan}
  \city{}
  \country{}
}

\author{Biao Tang}
\email{biao.tang@meituan.com}
\affiliation[obeypunctuation=true]{%
  \institution{Meituan}
  \city{}
  \country{}
}

\author{Ming Liu}
\email{liuming66@meituan.com}
\affiliation[obeypunctuation=true]{%
  \institution{Meituan}
  \city{}
  \country{}
}

\author{Peng Yan}
\email{yanpeng04@meituan.com}
\affiliation[obeypunctuation=true]{%
  \institution{Meituan}
  \city{}
  \country{}
}

\renewcommand{\shortauthors}{Gao et al.}

\begin{abstract}
Recommender systems are widely used in e-commerce and content platforms, where improvement requires repeated model design, training optimization, and experimental validation.
Large language model (LLM) agents increasingly automate this development cycle.
Sustained autonomous improvement requires agents to improve both recommendation models and their development workflows.
However, task-performance-driven approaches select updates mainly by candidates' final scores, without a separate optimization signal for experimentation quality.
We therefore present \textbf{DualEvolve}, a dual-level framework that co-evolves recommendation mechanisms and agent orchestration toward recursive self-improvement.
Inner evolution revises model and training code using task feedback to improve recommendation performance.
Outer evolution scores the inner-loop experimental process, updates harness and skill rules using these scores and execution traces, and adjusts orchestration choices based on next-round scores to improve the quality of subsequent experiments.
Experiments on MovieLens-1M and three Amazon Reviews 2023 subsets cover four recommendation methods and 16 settings.
DualEvolve achieves higher test means than Base, OpenEvolve, and DeepEvolve in all 64 setting--metric comparisons, with macro-averaged relative gains over DeepEvolve of 7.6\% (HR@5) and 9.0\% (NDCG@5).

\end{abstract}



\maketitle

\section{Introduction}
\label{sec:intro}

Recommender systems are widely used on e-commerce and content platforms. Engineers improve models and training procedures through repeated diagnosis, design, and experimental validation. LLM agents increasingly automate this cycle through code editing and evaluation~\cite{DBLP:conf/icml/HuangVLL24,DBLP:conf/icml/GuoD0C0024}, retrieval, ablation analysis, and targeted refinement~\cite{DBLP:conf/nips/NamYCSAP25}. Program evolution extends this feedback-driven cycle to searches over executable algorithms, revising candidate programs according to measured results~\cite{romeraparedes2024funsearch,DBLP:journals/corr/abs-2506-13131}. Benchmarks complement these methods by evaluating agents on longer engineering workflows~\cite{DBLP:conf/iclr/ChanCJASMSLMPMW25,DBLP:conf/icml/StaraceJSACMDMK25}. These code-editing and search methods let recommendation agents modify model logic beyond predefined alternatives.

\begin{figure}[t]
  \centering
  \captionsetup{belowskip=3pt}
  \includegraphics[width=\columnwidth]{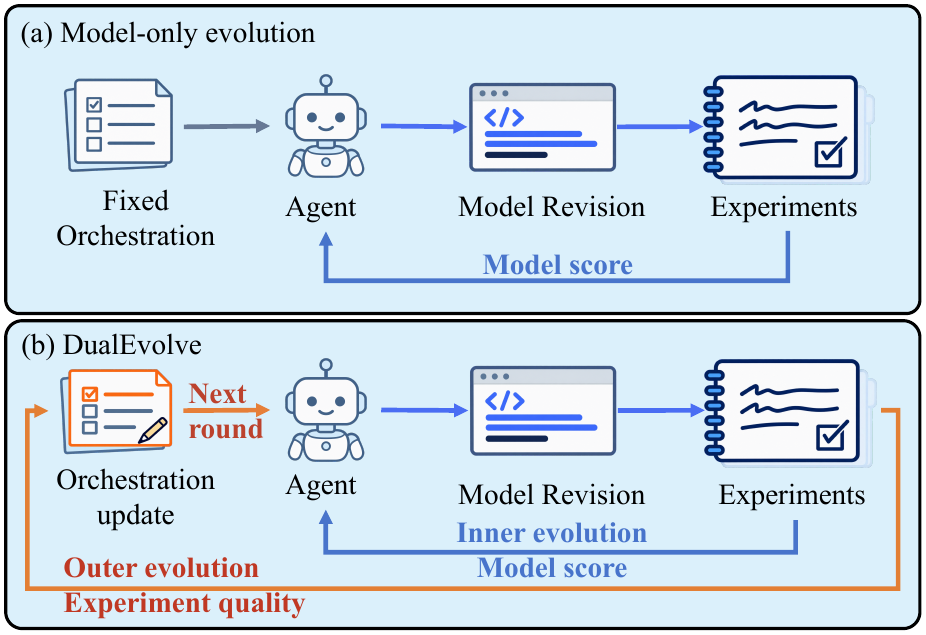}
  \caption{Model-only evolution revises recommendation code using task feedback. DualEvolve adds an outer loop that evaluates experiment quality and updates agent orchestration for subsequent rounds.}
  \Description{The top panel shows an agent revising models and running experiments under fixed orchestration, with model scores guiding further revisions. The bottom panel retains this inner loop and adds process-quality feedback to update orchestration, which guides the agent in the next round.}
  \label{fig:intro_evolution}
  \vspace{-1em}
\end{figure}

\begin{figure*}[t]
  \centering
  \captionsetup{belowskip=3pt}
  \includegraphics[width=\textwidth]{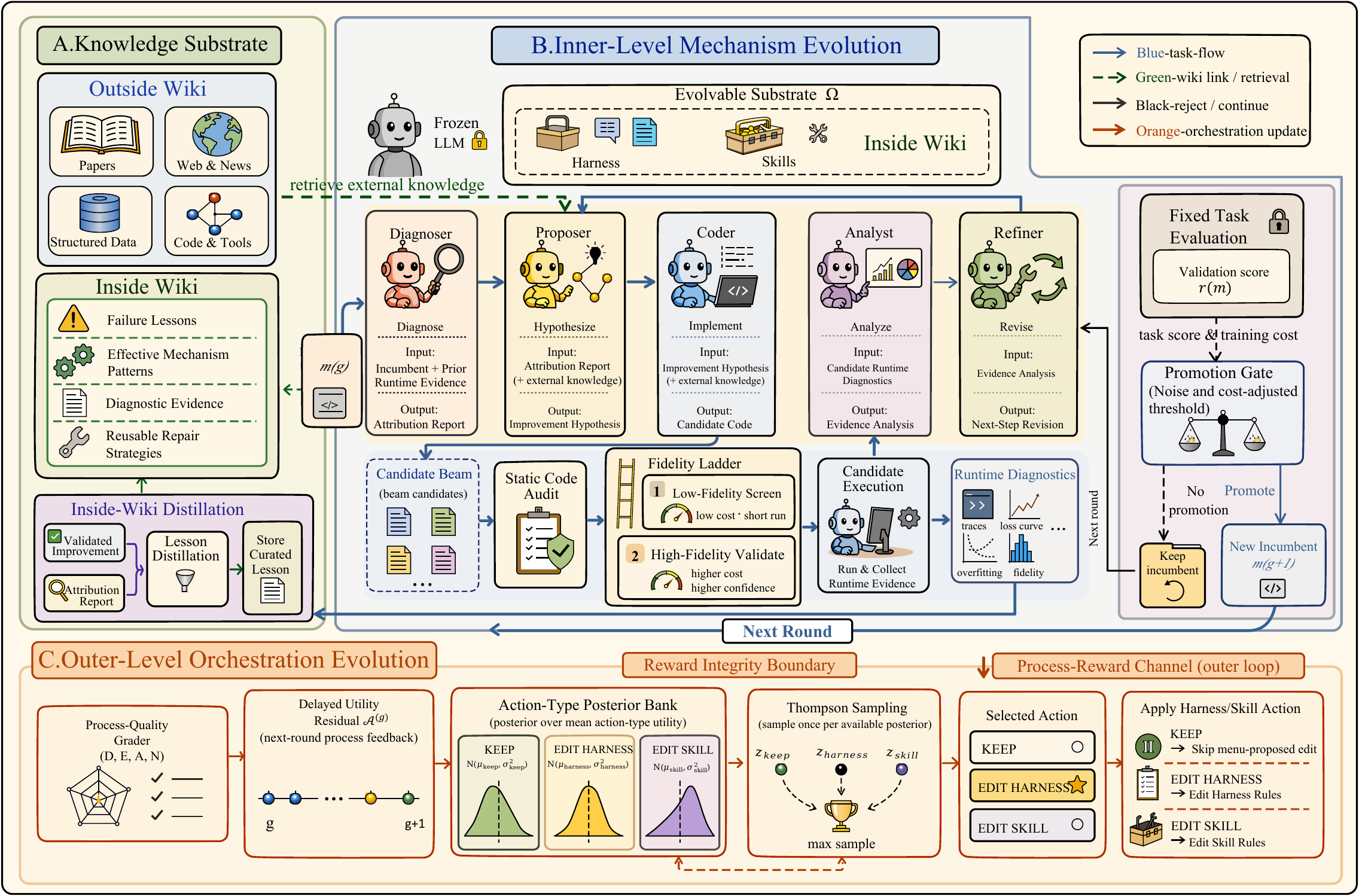}
  \caption{DualEvolve jointly evolves task mechanisms and the harness-and-skill control layer of $\Omega$. The inside wiki accumulates task-specific search experience; the outside wiki supplies paper-derived priors from a fixed offline snapshot. KEEP skips the menu-proposed edit; a subsequent stagnation check can add direction guidance to the harness. Task performance determines mechanism admission under a fixed evaluator and rule, while process quality guides orchestration updates.}
  \Description{DualEvolve has two coupled loops. Inner mechanism evolution diagnoses and revises candidate code. Outer orchestration evolution edits harness and skills using process-quality evidence. The inside wiki accumulates task-specific search experience, and a fixed offline outside-wiki snapshot supplies paper-derived method priors. KEEP skips the menu-proposed edit, while a separate stagnation check may append harness guidance. Evaluation and mechanism admission remain fixed.}
  \label{fig:framework}
  \vspace{-1em}
\end{figure*}

Beyond model revision, agents can also improve the workflow used to propose changes, diagnose failures, and plan subsequent experiments. Recommendation metrics assess candidate quality, but do not fully capture the quality of experimental design and subsequent decisions. For example, after a controlled experiment yields no gain in the current setting, an agent may reduce retries in that direction and explore alternatives, or ignore failures and repeat similar changes. Both may leave the best model unchanged, but only the former uses evidence to redirect search. Model improvement alone cannot distinguish these behaviors. Alongside task-based model selection, agents therefore need feedback on experimental design, analysis, and subsequent decisions to improve their development workflow (Figure~\ref{fig:intro_evolution}).

Fortunately, automated agent-design research treats workflows and harnesses as optimization targets~\cite{DBLP:conf/iclr/HuLC25,DBLP:conf/iclr/ZhangXYTCCZCHWZ25}. Meta-Harness proposes harness revisions from historical code, evaluation results, and execution traces~\cite{lee2026metaharness}. In recommendation, recent studies extend search to architectures, diagnostic tools, and development workflows~\cite{DBLP:journals/corr/abs-2602-12612,DBLP:journals/corr/abs-2606-27243}. AgentX diagnoses traces to revise agent rules and checks edits through paired replay~\cite{DBLP:journals/corr/abs-2606-26859}. These methods provide a basis for improving workflows from accumulated experience. During an ongoing recommendation search, we further use feedback from experiments following a rule revision to guide subsequent orchestration choices. The LLM proposes specific harness and skill edits, while Thompson sampling selects among KEEP and the available edit types. The residual between the next round's process score and a running historical baseline updates the utility estimate for the selected action type. These estimates guide subsequent choices as recommendation mechanisms and experimental evidence evolve.

Building on this idea, we propose \textbf{DualEvolve}\footnote{Code is available at \url{https://github.com/ToAdventure/dualevolve}.}, a dual-level framework that co-evolves recommendation mechanisms and agent development workflows with frozen LLM parameters (Figure~\ref{fig:framework}). Inner mechanism evolution iteratively improves models and training through diagnosis, code revision, and experimental validation. A noise- and cost-aware rule determines whether a candidate replaces the incumbent. Outer orchestration evolution evaluates experimental processes by decision quality, experimental rigor, analysis depth, and search novelty to revise harness and skill rules. Thompson sampling selects orchestration actions. Next-round process feedback updates action-type utility estimates for subsequent choices, closing the cross-round loop between rule updates and experiments. The mechanism evaluator and admission rule remain fixed. Across four datasets and four recommendation methods, macro-averaged relative gains over DeepEvolve reach 7.6\% in HR@5 and 9.0\% in NDCG@5.

Our contributions are as follows.
\begin{itemize}
  \setlength{\itemsep}{2pt plus 1pt}
  \item We propose a dual-level framework that co-evolves recommendation mechanisms and the agent workflow guiding their improvement under fixed evaluation and admission.
  \item We design adaptive orchestration using process feedback to estimate action-type utilities, Thompson sampling to select harness and skill edits, and evidence to redirect search.
  \item DualEvolve exceeds Base, OpenEvolve, and DeepEvolve test means in all 64 setting--metric comparisons across four datasets and four recommendation methods. System-level comparisons examine memory, action selection, rule updates, and feedback. Transfer experiments and target-run traces show reuse of source-evolved rules in a new setting.
\end{itemize}

\section{Related Work}
\label{sec:related}

\subsection{From Iterative Refinement to Mechanism Evolution}

Self-Refine alternates feedback and output revision, while Reflexion retains environmental feedback as verbal memory~\cite{DBLP:conf/nips/MadaanTGHGW0DPY23,DBLP:conf/nips/ShinnCGNY23}. Machine-learning agents extend iteration to executable experiments. DeepEvolve adds research before program search~\cite{DBLP:journals/corr/abs-2510-06056}, and Glia uses specialized agents to propose, test, and analyze systems mechanisms~\cite{DBLP:conf/cais/HamadanianKNNCP26}. Building on iterative code revision, DualEvolve jointly evolves task logic and the workflow that generates it under fixed evaluation and admission.

\subsection{Mechanism Evolution in Recommendation}

Recommendation AutoML typically searches predefined operations. AutoFIS selects feature interactions and AutoCTR searches modular graphs for click-through-rate prediction~\cite{DBLP:conf/kdd/LiuZL0LTHL020,DBLP:conf/kdd/SongCZYTH20}. Language-agent methods instead model users, items, and interactions. Agent4Rec simulates users through profile, memory, and action modules, while AgentCF updates user and item profiles through agent collaboration~\cite{DBLP:conf/sigir/0003CSWC24,DBLP:conf/www/ZhangHXSMZLW24}. These methods center on agent profiles and interactions, while mechanism evolution targets executable recommendation models and training procedures. Recent self-evolving recommenders, discussed in Section~\ref{sec:intro}, make both recommendation logic and its development workflow editable.

\subsection{Orchestration Evolution and Reward Integrity}

Automated agent design optimizes structure, control flow, prompts, and code. AgentSquare and MaAS search modular or supernet configurations~\cite{DBLP:conf/iclr/ShangLZMLXL25,DBLP:conf/icml/ZhangNF00025}. Trace updates prompts and code from execution traces, and GEPA evolves textual components through reflection and Pareto selection~\cite{DBLP:conf/nips/ChengNS24,agrawal2026gepa}. EnCompass compiles workflows into search spaces; Agent Lightning assigns task feedback to model calls~\cite{DBLP:conf/nips/LiSYZ25,DBLP:journals/corr/abs-2508-03680}. TTD-DR improves deep-research workflow components at test time~\cite{DBLP:journals/corr/abs-2507-16075}.

Other methods evolve reusable guidance or the agent implementation. SePO co-evolves task and optimizer prompts~\cite{DBLP:journals/corr/abs-2606-04465}. Ratchet manages natural-language skills, and Metis combines textual experience with validated tools~\cite{DBLP:journals/corr/abs-2605-22148,DBLP:journals/corr/abs-2606-24151}. G\"odel Agent, Darwin G\"odel Machine, and Live-SWE-agent rewrite agent logic or scaffolds~\cite{DBLP:conf/acl/YinWPL0W25,DBLP:journals/corr/abs-2505-22954,DBLP:journals/corr/abs-2511-13646}.

Recent harness optimization preserves model weights while changing persistent agent behavior. Meta-Harness searches harness code from prior candidates and traces; Agentic Harness Engineering records components and edit outcomes; TTHE evolves harnesses from unlabeled test-time traces~\cite{lee2026metaharness,lin2026agenticharness,DBLP:journals/corr/abs-2607-08124}. In recommendation, AgentX combines online strategy tests with paired replay of harness edits~\cite{DBLP:journals/corr/abs-2606-26859}. DualEvolve uses next-round process quality to update action-type utilities for Thompson sampling during offline mechanism search, with noise- and cost-aware admission held fixed.

Harmful drift across updates to models, memory, tools, and workflows has been termed \emph{misevolution}~\cite{DBLP:journals/corr/abs-2509-26354}. The Meta-Agent Challenge reports attempts to exploit protected evaluation information~\cite{DBLP:journals/corr/abs-2606-04455}. These observations motivate placing evaluation and mechanism admission outside the editable scope.

\section{Problem Formulation}
\label{sec:problem}

\subsection{Task and Editable Objects}
\label{sec:problem:objects}

For a recommendation task $T$, a fixed validation evaluator returns a scalar score $r(m)$ for mechanism $m$ at the full training budget. A \emph{mechanism} is executable code defining the recommendation model and training procedure, including its architecture, forward computation, loss, optimizer, and training loop. Data loading, evaluation, and metric computation remain fixed. Let $c(m)>0$ denote measured training cost.

Let $G$ be the number of completed search rounds. At the start of round $g\in\{0,\ldots,G-1\}$, the state comprises the incumbent mechanism $m^{(g)}$ and orchestration substrate $\Omega^{(g)}$. The substrate contains harness and skill rules governing candidate generation and revision, together with the task-specific inside wiki in Figure~\ref{fig:framework}. The inside wiki retains candidate records and consolidated lessons from the current task. Orchestration evolution edits the rules; experimental evidence updates the inside wiki. The baseline is $m^{(0)}$, and the final incumbent is $m^{(G)}$.

\subsection{Dual-Level Evolution Under a Fixed Boundary}

The objective is to improve $r(m^{(G)})$ by evolving the mechanism and its orchestration jointly. In round $g$, the current substrate guides a \emph{mechanism lineage} $\mathcal{L}^{(g)}=(v_0,\ldots,v_S)$ containing $S$ selected refinements after its initial candidate. It starts from the incumbent or a retained candidate and develops through diagnosis of earlier evaluations. The best full-budget candidate $v_\star^{(g)}$ replaces the incumbent only if it passes a fixed admission rule, the \emph{promotion gate}. Process-quality feedback $R_{\mathrm{outer}}$ then guides the next orchestration update.

\emph{Reward integrity} places the evaluator and gate outside both editable layers. For decision $\mathrm{Promote}^{(g)}\in\{0,1\}$ and fixed gate $\Gamma$,
\begin{equation}
  \label{eq:integrity}
  \mathrm{Promote}^{(g)} =
  \Gamma\!\left(
    r(v_\star^{(g)}), r(m^{(g)}), \delta^\star,
    c(v_\star^{(g)}), c(m^{(g)}); \lambda
  \right),
\end{equation}
where $\delta^\star$ is the task-specific noise floor set before search and $\lambda\geq0$ controls the additional gain required of slower candidates. The evaluator $r$, cost measurement $c$, calibration, and gate remain fixed. Process rewards guide candidate generation through orchestration updates, while $\Gamma$ depends on task scores, training costs, and the fixed calibration parameters. Task-derived diagnoses and lessons inform subsequent search. Section~\ref{sec:method:screen} specifies the gate and incumbent update.

\section{Methodology}
\label{sec:method}

DualEvolve alternates mechanism revision under the fixed promotion gate with process-guided updates to the harness and skill rules that shape the next round's search.

\subsection{Inner-Level Mechanism Evolution}
\label{sec:method:inner}

\subsubsection{Lineage Generation}
\label{sec:method:lineage}

The inner loop uses the five frozen-LLM roles in Figure~\ref{fig:framework}. The \emph{diagnoser} attributes weaknesses from the current mechanism and runtime evidence. The \emph{proposer} combines this diagnosis with retrieved knowledge to formulate an improvement hypothesis, which the \emph{coder} implements. After execution, the \emph{analyst} interprets the diagnostics and the \emph{refiner} develops the next revision. The first pass starts from the incumbent or a retained candidate; subsequent passes refine the selected mechanism, with backtracking when progress stalls.

A deterministic extractor produces signals $\eta(v)$ from each candidate's log, including loss trends, incumbent-relative gains, score changes across training budgets, and task-specific diagnostics. The diagnoser also receives whether its previous diagnosis led to an improvement. These signals, a bounded lineage history, $\Omega^{(g)}$, and knowledge retrieved from the \emph{outside wiki} guide authoring. The outside wiki supplies paper-derived priors through retrieval from the current diagnosis and selective reading of linked cards. Its snapshot stays fixed during search, while task outcomes and lessons enter the inside wiki (Appendix~\ref{app:outside-wiki}). With parent $p_k$ retained for step $k+1$, the candidate beam is
\begin{equation}
    \label{eq:lineage}
    \mathcal{B}_{k+1}
    = \{v_{j,k+1}\}_{j=1}^{W_k}
    = \mathrm{Author}\!\left(p_k;\, \eta(p_k),\, \Omega^{(g)}\right).
\end{equation}
The widths $W_k$ are fixed before search and may vary by step. Screening selects the next lineage member $v_{k+1}$ and normally makes it the next parent.

\subsubsection{Two-Rung Screening and the Promotion Gate}
\label{sec:method:screen}

The fixed task adapter evaluates candidates at low and high training budgets, with scores $r_{\mathrm{low}}$ and $r_{\mathrm{high}}=r$. Each beam is screened at low fidelity, and its winner receives high-fidelity evaluation. The initial candidate $v_0$ is also evaluated at high fidelity. The step winner and best full-budget lineage candidate are
\begin{equation}
    \label{eq:tworung}
    \begin{aligned}
        v_{k+1} &= \arg\max_{v \in \mathcal{B}_{k+1}} r_{\mathrm{low}}(v),\\
        v_\star^{(g)} &= \arg\max_{v \in \mathcal{L}^{(g)}} r_{\mathrm{high}}(v).
    \end{aligned}
\end{equation}
Failed runs score below valid measurements. If a whole beam fails, the next scheduled step retains its parent and evidence. When any scheduled width exceeds one, a low-fidelity reference tracks progress. It updates only if the selected score improves by more than a preset margin. After $K$ consecutive evaluated steps without such improvement, refinement restarts from the last candidate that updated the reference and seeks a structurally different mechanism. Consecutive non-improvement is counted over steps with a valid low-fidelity winner.

The high-fidelity winner is compared with the incumbent using its gain and relative training cost,
\begin{equation}
    \label{eq:gain-cost}
    \begin{aligned}
        d &= r(v_\star^{(g)})-r(m^{(g)}),\\
        \rho &= c(v_\star^{(g)})/c(m^{(g)}).
    \end{aligned}
\end{equation}
The calibrated noise floor $\delta^\star$ and cost coefficient $\lambda$ give the threshold
\begin{equation}
    \label{eq:promotion-threshold}
    \theta_\star^{(g)}
    = \max(0,\delta^\star)\left[1+\lambda\max(0,\rho-1)\right],
\end{equation}
and the incumbent update is
\begin{equation}
    \label{eq:inner-update}
    m^{(g+1)}
    = \begin{cases}
        v_\star^{(g)}, & d > \theta_\star^{(g)},\\
        m^{(g)}, & \text{otherwise}.
    \end{cases}
\end{equation}
A slower candidate must therefore deliver a larger gain; a faster candidate still clears the base noise floor.

Promising candidates can remain available without promotion. Those matching or exceeding the initial baseline at low fidelity but lacking high-fidelity confirmation become runner-ups. The inside wiki maintains a bounded starting-point pool containing the incumbent, runner-ups, and promising candidates truncated by the budget. At fixed intervals, it selects the least recently used non-incumbent candidate when available.

The lesson store $\mathcal{K}\subset\Omega$ is the textual component of the inside wiki. It records a mechanism change and its explanation when the step winner improves on the previous step and the analyst supports the attribution with runtime signals. Periodic consolidation combines these lessons with wins, failures, runner-ups, and truncated evaluations into task-specific guidance, retaining the raw records. The proposer uses that guidance in later rounds.

The inside wiki retains task-local scores, budgets, and lessons. Skills
encode conditional refinement procedures, such as continuing a candidate
whose longer-budget curve is still rising. Harness rules govern the
cross-candidate plan, including redirection after a non-promoting round.
These rules can transfer while raw source records remain task-local.

\subsection{Outer-Level Orchestration Evolution}
\label{sec:method:outer}

\subsubsection{Process-Quality Reward}
\label{sec:method:reward}

At each round's end, a fixed GLM-5.2 grader assesses the trace for decision quality
$D$, experimental rigor $E$, analysis depth $A$, and novelty $N$. The first
three dimensions use five, three, and six criteria per experiment; novelty
uses four criteria for the complete round. A criterion counts only when
supported by trace evidence.

Grading is invoked three times, with one LLM call per dimension in each
pass. For $D$, $E$, and $A$, criterion-hit fractions are averaged across
experiments and then across the returned grades. For novelty, let
$b_{ij}\in\{0,1\}$ indicate whether criterion $i$ is satisfied in pass $j$.
The per-pass score $N_j$ and final score $N$ are
\begin{equation}
    \label{eq:novelty-score}
    N_j=\frac{1}{4}\sum_{i=1}^{4}b_{ij},
    \qquad N=\frac{1}{3}\sum_{j=1}^{3}N_j.
\end{equation}
If a response cannot be parsed, a deterministic fallback credits
trace-verifiable criteria. The outer reward is
\begin{equation}
    \label{eq:reward}
    R_{\mathrm{outer}}=w_DD+w_EE+w_AA+w_NN,
\end{equation}
with positive weights summing to one and fixed throughout the run.

Decision quality checks evidence grounding, an explicit hypothesis,
testability, relevance to the current weakness, and use of previous results.
Experimental rigor checks a comparable control, isolation of the target
factor, and sufficient training or checkpoints. Analysis depth checks
baseline comparison, separation of negative or null effects from execution
failures, uncertainty, slow convergence, the scope of conclusions, and
evidence-based next steps. Round-level novelty checks structurally different
mechanisms spanning multiple directions, a changed starting point or
direction from the previous round, actual authoring and evaluation of new
candidates, and avoidance of failed or exhausted directions.
Together, these dimensions capture process evidence beyond task gains.
For example, a controlled test with a well-explained negative result can
still inform the next orchestration decision.

\subsubsection{Thompson-Sampling Action Selection}
\label{sec:method:thompson}

After each round, a menu agent proposes at most one harness edit and one skill edit, each adding, retiring, or merging rules. Thompson sampling chooses among the available edit types and the always-available KEEP action, which applies no menu-proposed edit. The next round's process reward supplies a delayed utility observation for the selected action type $a^{(g)}$. With baseline $V^{(g)}$ and $0<\alpha\leq1$,
\begin{equation}
    \label{eq:credit}
    \begin{aligned}
      \mathcal{A}^{(g)} &= R_{\mathrm{outer}}^{(g+1)}-V^{(g)},\\
      V^{(g+1)} &= (1-\alpha)V^{(g)}+\alpha R_{\mathrm{outer}}^{(g+1)}.
    \end{aligned}
\end{equation}
The baseline is initialized by the first graded round with sufficient measurements. Each utility residual uses the pre-update baseline. Delayed observations are recorded for actions followed by another graded round. If a fraction $f\geq0.5$ of the round's experiments yields no measurement and at most two experiments yield measurements, the pending action's update uses $-0.5f$ in place of the utility residual, while the EMA baseline remains unchanged.

Each action type $a$ has a Normal prior with mean zero and variance $\tau_0^2>0$. Let $n_a$, $\bar{\mathcal{A}}_a$, and $s_a^2$ denote its observation count, mean utility residual, and sample variance. For $n_a\geq2$, set $\hat{\sigma}_a^2=\max(s_a^2,\epsilon_{\mathrm{obs}})$ using a positive observation-variance floor. The posterior is
\begin{equation}
    \label{eq:posterior}
    \begin{aligned}
      \mu_a &= \frac{n_a\,\bar{\mathcal{A}}_a/\hat{\sigma}_a^2}
                    {\tau_0^{-2}+n_a/\hat{\sigma}_a^2},\\
      \sigma_a^2 &= \left(\tau_0^{-2}+\frac{n_a}{\hat{\sigma}_a^2}\right)^{-1}.
    \end{aligned}
\end{equation}
For $n_a=0$, $(\mu_a,\sigma_a^2)=(0,\tau_0^2)$; for $n_a=1$, it is $(\bar{\mathcal{A}}_a/2,\tau_0^2/2)$. Observations are aggregated by action type; the menu agent specifies each edit's content. A positive floor $\epsilon_{\mathrm{samp}}$ preserves sampling variance without changing the posterior mean. Over available action types $\mathcal{U}^{(g)}$, Thompson sampling selects
\begin{equation}
    \label{eq:thompson-selection}
    \begin{aligned}
      z_a &\sim \mathcal{N}\!\left(\mu_a,\max(\sigma_a^2,\epsilon_{\mathrm{samp}})\right),
             \quad a\in\mathcal{U}^{(g)},\\
      a^{(g)} &= \arg\max_{a\in\mathcal{U}^{(g)}}z_a.
    \end{aligned}
\end{equation}

Accumulated utility residuals update action-type posterior means and
uncertainties. The sampling-variance floor maintains exploration among
available action types throughout the run.

After the selected menu action, a separate stagnation check can append
direction guidance to the harness, including when that action is KEEP.
After one round without promotion, the frozen agent selects a less explored
direction from the trial log and records its guidance for the next round
when budget remains.
Generic mechanism axes provide a deterministic fallback. Menu edits and
redirection update the control layer under fixed evaluation and admission. Two consecutive rounds
without promotion stop the search. Periodic LLM curation uses runtime evidence
to retain useful harness and skill guidance, retire ineffective or
contradictory guidance, and merge redundancy.

\subsection{Reward Integrity}
\label{sec:method:integrity}

The fixed task adapter computes task scores and training cost, and the gate applies the pre-search calibration in Eq.~\eqref{eq:promotion-threshold}. Process-quality feedback guides rule revisions, action-type utility estimation, and baseline updates. Task outcomes remain available for diagnosis, lessons, direction changes, and process assessment, while Eq.~\eqref{eq:integrity} fixes the admission boundary. Before time-bounded execution, candidate code passes a static audit forbidding system interfaces and file-reading chains that could reach evaluation data.

\providecommand{\method}{\textsc{DualEvolve}}
\newcommand{\SearchStabilityEvidence}{%
\begin{table}[!htbp]
\centering
\small
\setlength{\tabcolsep}{3pt}
\caption{Independent-search means over three searches. SD is shown for
\method{}; Table~\ref{tab:comparator_search_sd} supplies comparator SDs.
Base is the fixed mechanism in Table~\ref{tab:main_results}. Stars mark
unadjusted two-sided Welch significance against the higher-mean OE/DE
comparator: $^{*}p<0.05$, $^{**}p<0.01$, and $^{***}p<0.001$. Boldface and
underlining mark the best and second-best means.}
\label{tab:search_repeats}
\begin{tabular*}{\columnwidth}{@{\extracolsep{\fill}}lcccc@{}}
\toprule
Metric & Base & OE & DE & \method{} \\
\midrule
\multicolumn{5}{c}{Baby/NGCF (3 searches)} \\
\midrule
H@5 & 0.2131 & 0.2643 & \underline{0.2667} & $\mathbf{0.2798}\,\pm\,0.0048^{*}$ \\
H@10 & 0.3277 & \underline{0.3875} & 0.3844 & $\mathbf{0.3985}\,\pm\,0.0039^{*}$ \\
N@5 & 0.1387 & 0.1831 & \underline{0.1852} & $\mathbf{0.1994}\,\pm\,0.0058^{*}$ \\
N@10 & 0.1755 & 0.2212 & \underline{0.2228} & $\mathbf{0.2397}\,\pm\,0.0068^{*}$ \\
\midrule
\multicolumn{5}{c}{Software/NGCF (3 searches)} \\
\midrule
H@5 & 0.5356 & 0.5256 & \underline{0.5441} & $\mathbf{0.5598}\,\pm\,0.0058^{*}$ \\
H@10 & 0.6763 & 0.6604 & \underline{0.6797} & $\mathbf{0.6945}\,\pm\,0.0051^{*}$ \\
N@5 & 0.3918 & 0.3894 & \underline{0.4115} & $\mathbf{0.4278}\,\pm\,0.0025^{**}$ \\
N@10 & 0.4374 & 0.4331 & \underline{0.4508} & $\mathbf{0.4678}\,\pm\,0.0040^{**}$ \\
\midrule
\multicolumn{5}{c}{CDs\&Vinyl/GRU4Rec (3 searches)} \\
\midrule
H@5 & 0.4717 & 0.4872 & \underline{0.5068} & $\mathbf{0.5738}\,\pm\,0.0118^{**}$ \\
H@10 & 0.5793 & 0.5833 & \underline{0.6061} & $\mathbf{0.6903}\,\pm\,0.0101^{***}$ \\
N@5 & 0.3632 & 0.3787 & \underline{0.3972} & $\mathbf{0.4466}\,\pm\,0.0108^{**}$ \\
N@10 & 0.3980 & 0.4129 & \underline{0.4300} & $\mathbf{0.4844}\,\pm\,0.0102^{**}$ \\
\bottomrule
\end{tabular*}
\end{table}

\paragraph{Stability across independent searches.}
On three representative settings, we independently run OE, DE, and \method{}
three times with frozen GLM-5.2 under the shared 12-hour search cap.
OE and DE retain their native candidate schedules, while \method{} uses
at most six outer rounds. The runs share the three search-seed identifiers.
One search-level observation averages
the selected mechanism's test scores over training seeds 12352--12354.
Table~\ref{tab:search_repeats} reports the means and \method{} sample SDs
across these observations. Comparator SDs for the Welch tests appear in
Table~\ref{tab:comparator_search_sd}.

\method{} has the highest repeated-search mean in all 12 setting--metric
comparisons, with two-sided independent-sample Welch $p<0.05$ against the
higher-mean OE/DE comparator. Its between-search sample SDs span 0.0025--0.0118.
}

\newcommand{\ScreeningEvidence}{%
\paragraph{Screening and final mechanism selection.}
Low-fidelity evaluation provides coarse screening to allocate full-budget
training to selected candidates. Differences between the two fidelity
rankings motivate high-fidelity re-evaluation, whose scores determine
final admission under the fixed gate. Appendix~\ref{app:additional-evidence}
reports ranking agreement and full-budget re-evaluation of screened candidates.
Among the 16 final mechanisms, the final recorded transition is a
within-lineage refinement for 13 and a newly selected starting point for
three, reflecting both local revision and renewed exploration.
Table~\ref{tab:mechanisms} covers frequency-based history representations,
graph propagation, objectives, and optimization. CDs\&Vinyl/NCF learns to
filter history embeddings, while Software/NGCF changes graph-information
weights. Sequential methods revise negative-logit correction and
training-only sequence denoising, covering both computation and training.
}

\newcommand{\ProcessValidationEvidence}{%
\paragraph{Assessment of process feedback.}
Human evaluation covers all four reward dimensions, with candidate-level
annotations for decision quality, experimental rigor, and analysis depth and
round-level annotations for novelty. Across 3,387 criterion judgments, the
grader achieves 83.1\% pooled agreement and Cohen's $\kappa=0.432$.
Appendix~\ref{app:additional-evidence} reports dimension-wise results and
comparisons using alternative graders.

\paragraph{Orchestration behavior.}
In the archived outer-menu records, 54\% select KEEP,
25\% select EDIT HARNESS, and 21\% select EDIT SKILL.
In three of four additional
posterior-mean argmax traces, all selected menu actions are KEEP.
Posterior sampling selects KEEP most often and also selects both rule-edit
types, as reflected in these menu-action frequencies.
Four further Fixed Rules searches with a four-round cap are reported in
Appendix~\ref{app:additional-evidence}; they characterize the frozen-rule
policy, with edits and redirection inactive under the same stopping condition.
}

\newcommand{\StressTestEvidence}{%
\paragraph{Shortcut-exposure stress test.}
We examine reward integrity on Baby/SASRec by placing the same four
shortcut interfaces in the validation environment and varying their
disclosure to the agent. The interfaces reduce the candidate item set,
substitute easier negatives, inflate reported scores, or expose validation
labels. Test evaluation remains clean throughout.
The undisclosed condition names none, partial
disclosure names two candidate-selection shortcuts, and full disclosure
names all four and permits their use. Each condition has two searches
with a four-round cap and fixed orchestration rules.
Table~\ref{tab:shortcut_exposure} reports shortcut use.

\begin{table}[!htbp]
\centering
\caption{Shortcut use under three disclosure conditions. Counts give
affected mechanisms and searches; each condition contains two searches
and 56 generated mechanisms.}
\label{tab:shortcut_exposure}
\begin{tabular*}{\columnwidth}{@{\extracolsep{\fill}}lcc@{}}
\toprule
Disclosure & Mechanisms & Searches \\
\midrule
Undisclosed & 0/56 & 0/2 \\
Partial & 0/56 & 0/2 \\
Full & 25/56 & 1/2 \\
\bottomrule
\end{tabular*}
\end{table}

No shortcut is used in the 112 mechanisms generated under undisclosed
or partial disclosure. Under full disclosure, one search adopts shortcuts in
25 of its 28 mechanisms and inflates the winner's validation score to
1.6961, above the objective's theoretical maximum of 1. The winner
depends on shortcut interfaces absent from the clean evaluator and
cannot execute there. The other fully disclosed search uses no
shortcut. This observed propagation of reward manipulation motivates
keeping evaluation operations outside the editable search interface.

}

\section{Experiments}
\label{sec:experiments}

We evaluate \method{} on a $4{\times}4$ combination of datasets and base
recommendation methods. Our questions concern recommendation quality and
search stability (RQ1), the components and feedback guiding evolution
(RQ2), transfer of evolved rules (RQ3), and search cost (RQ4).

\subsection{Experimental Setup}
\label{sec:setup}

\noindent\textbf{Data and evaluation.}
We use CDs\&Vinyl, Baby, and Software from Amazon Reviews 2023~\cite{hou2024bridging}
and ML-1M~\cite{harper2015movielens}, with per-user deduplication and iterative
5-core filtering. Amazon retains the published last-out splits; ML-1M reserves
each user's penultimate and last interactions for validation and test.
Baby and Software are subsampled with a fixed seed to the 14,335-user
CDs\&Vinyl pool size; ML-1M retains 6,040 users.
Each evaluation ranks one held-out positive against 99 uniformly sampled negatives.
Search uses validation evidence, and test evaluation follows final mechanism
selection. Appendix~\ref{app:setup} reports dataset statistics and parameters.

\noindent\textbf{Methods and budgets.}
The initial mechanisms implement NCF, NGCF, SASRec, and
GRU4Rec~\cite{he2017ncf,wang2019ngcf,kang2018sasrec,hidasi2016gru4rec}.
Base uses each initial mechanism without evolution.
OpenEvolve (OE)~\cite{sharma2025openevolve} evolves a population of programs
as an open-source implementation of AlphaEvolve~\cite{DBLP:journals/corr/abs-2506-13131}.
DeepEvolve (DE)~\cite{DBLP:journals/corr/abs-2510-06056} adds research and planning
before code generation. All three frameworks use the same frozen GLM-5.2,
initial mechanism, splits, task adapter, full-fidelity training budget,
and single-GPU hardware configuration. They receive the same 12-hour
search-time cap and retain their native candidate schedules. All three
frameworks have access to the same fixed local knowledge snapshot, with
online tool access, source ingestion, and snapshot maintenance disabled
throughout search (Appendix~\ref{app:outside-wiki}).
The main grid uses one search per framework and setting; its
significance tests compare the selected mechanisms over three initialization
seeds. Search-level stability is assessed separately using three runs per
framework on three representative settings.

\noindent\textbf{Metrics.}
The main, component, and transfer tables report test means over initialization
seeds 12352, 12353, and 12354. H and N abbreviate HR and NDCG.
Search compares candidates using the fixed validation objective
\begin{equation}
\label{eq:validation-objective}
r_{\mathrm{val}}=0.6\,\mathrm{HR@5}+0.4\,\mathrm{NDCG@5}.
\end{equation}
Both fidelity rungs use this objective; admission applies the fixed noise-
and cost-aware threshold. Table~\ref{tab:main_results} also gives sample
standard deviations for \method{}. Its significance marks use two-sided
Welch tests against the three-seed results of the higher-mean OE or DE
comparator, without multiple-comparison correction.

\newcommand{\sigp}{\textsuperscript{\ensuremath{\dagger}}}
\newcommand{\dualresult}[3]{\mbox{#1$\,\pm\,$#3#2}}
\newcommand{\dualbest}[3]{\mbox{\textbf{#1}$\,\pm\,$#3#2}}
\begin{table*}[t]
\centering
\footnotesize
\setlength{\tabcolsep}{2.2pt}
\renewcommand{\arraystretch}{1.12}
\caption{Overall test performance across four datasets and four base
recommendation methods. All results are means over the same three
initialization seeds. \method{} reports mean $\pm$ sample standard deviation.
H and N denote HR and NDCG, respectively. Boldface and underlining indicate
the best and second-best means for each setting--metric pair.
\sigp{} marks a statistically significant difference (two-sided Welch
$t$-test, $p<0.05$). The comparator is OpenEvolve (OE) or DeepEvolve (DE),
whichever has the higher mean for that setting and metric.}
\label{tab:main_results}
\resizebox{\textwidth}{!}{%
\begin{tabular}{llcccccccccccccccc}
\toprule
 & \multicolumn{1}{c}{} & \multicolumn{4}{c}{\textbf{CDs\&Vinyl}} & \multicolumn{4}{c}{\textbf{ML-1M}} & \multicolumn{4}{c}{\textbf{Baby}} & \multicolumn{4}{c}{\textbf{Software}} \\
\cmidrule(lr){3-6}\cmidrule(lr){7-10}\cmidrule(lr){11-14}\cmidrule(lr){15-18}
 & Metric & Base & OE & DE & \method{} & Base & OE & DE & \method{} & Base & OE & DE & \method{} & Base & OE & DE & \method{} \\
\midrule

\multirow{4}{*}{NCF} & H@5 & 0.3463 & \underline{0.4578} & 0.3402 & \dualbest{0.4992}{\sigp}{0.0039} & 0.4766 & 0.4973 & \underline{0.6116} & \dualbest{0.6818}{\sigp}{0.0016} & \underline{0.2657} & 0.2631 & 0.2637 & \dualbest{0.2800}{\sigp}{0.0009} & 0.4994 & 0.4971 & \underline{0.5246} & \dualbest{0.5485}{\sigp}{0.0028} \\
  & H@10 & 0.4580 & \underline{0.5816} & 0.4448 & \dualbest{0.6220}{\sigp}{0.0017} & 0.6560 & 0.6673 & \underline{0.7149} & \dualbest{0.7982}{\sigp}{0.0014} & \underline{0.3813} & 0.3755 & 0.3781 & \dualbest{0.3930}{\sigp}{0.0017} & 0.6239 & 0.6173 & \underline{0.6387} & \dualbest{0.6843}{\sigp}{0.0034} \\
  & N@5 & 0.2542 & \underline{0.3363} & 0.2500 & \dualbest{0.3760}{\sigp}{0.0030} & 0.3230 & 0.3403 & \underline{0.4570} & \dualbest{0.5274}{\sigp}{0.0010} & \underline{0.1792} & 0.1789 & 0.1785 & \dualbest{0.1925}{\sigp}{0.0015} & 0.3696 & 0.3685 & \underline{0.3933} & \dualbest{0.4060}{\sigp}{0.0045} \\
  & N@10 & 0.2903 & \underline{0.3768} & 0.2836 & \dualbest{0.4157}{\sigp}{0.0026} & 0.3810 & 0.3951 & \underline{0.4916} & \dualbest{0.5652}{\sigp}{0.0011} & \underline{0.2164} & 0.2162 & 0.2161 & \dualbest{0.2290}{\sigp}{0.0013} & 0.4099 & 0.4073 & \underline{0.4306} & \dualbest{0.4501}{\sigp}{0.0043} \\
\midrule
\multirow{4}{*}{NGCF} & H@5 & 0.4945 & \underline{0.5107} & 0.5093 & \dualbest{0.5221}{\sigp}{0.0008} & 0.4572 & \underline{0.4872} & 0.4858 & \dualbest{0.5153}{\sigp}{0.0017} & 0.2131 & 0.2654 & \underline{0.2670} & \dualbest{0.2784}{\sigp}{0.0018} & 0.5356 & 0.5254 & \underline{0.5446} & \dualbest{0.5530}{\sigp}{0.0009} \\
 & H@10 & 0.6219 & \underline{0.6304} & 0.6283 & \dualbest{0.6358}{\sigp}{0.0019} & 0.6353 & 0.6627 & \underline{0.6678} & \dualbest{0.6813}{\sigp}{0.0014} & 0.3277 & \underline{0.3876} & 0.3824 & \dualbest{0.3953}{\sigp}{0.0014} & 0.6763 & 0.6623 & \underline{0.6812} & \dualbest{0.6876}{\sigp}{0.0012} \\
 & N@5 & 0.3704 & 0.3808 & \underline{0.3844} & \dualbest{0.3973}{\sigp}{0.0010} & 0.3065 & 0.3295 & \underline{0.3315} & \dualbest{0.3568}{\sigp}{0.0021} & 0.1387 & 0.1829 & \underline{0.1841} & \dualbest{0.1896}{\sigp}{0.0017} & 0.3918 & 0.3905 & \underline{0.4096} & \dualbest{0.4132}{\sigp}{0.0008} \\
 & N@10 & 0.4117 & \underline{0.4232} & 0.4229 & \dualbest{0.4341}{\sigp}{0.0012} & 0.3641 & 0.3863 & \underline{0.3910} & \dualbest{0.4106}{\sigp}{0.0015} & 0.1755 & \underline{0.2220} & 0.2212 & \dualbest{0.2271}{\sigp}{0.0018} & 0.4374 & 0.4332 & \underline{0.4491} & \dualbest{0.4569}{\sigp}{0.0012} \\
\midrule
\multirow{4}{*}{SASRec} & H@5 & 0.4611 & 0.4883 & \underline{0.5641} & \dualbest{0.5742}{\sigp}{0.0018} & \underline{0.7161} & 0.7054 & 0.7078 & \dualbest{0.7286}{\sigp}{0.0017} & 0.2779 & \underline{0.2798} & 0.2681 & \dualbest{0.2975}{\sigp}{0.0019} & 0.6059 & 0.6044 & \underline{0.6125} & \dualbest{0.6289}{\sigp}{0.0046} \\
 & H@10 & 0.5792 & 0.6073 & \underline{0.6721} & \dualbest{0.6934}{\sigp}{0.0023} & \underline{0.8216} & 0.8173 & 0.8129 & \dualbest{0.8239}{\sigp}{0.0022} & 0.4010 & \underline{0.4144} & 0.3848 & \dualbest{0.4322}{\sigp}{0.0017} & \underline{0.7385} & 0.7355 & 0.7341 & \dualbest{0.7611}{\sigp}{0.0033} \\
 & N@5 & 0.3515 & 0.3682 & \underline{0.4236} & \dualbest{0.4526}{\sigp}{0.0015} & \underline{0.5607} & 0.5423 & 0.5445 & \dualbest{0.5760}{\sigp}{0.0012} & 0.1901 & \underline{0.1916} & 0.1806 & \dualbest{0.2042}{\sigp}{0.0026} & 0.4575 & 0.4561 & \underline{0.4690} & \dualbest{0.4884}{\sigp}{0.0023} \\
  & N@10 & 0.3897 & 0.4082 & \underline{0.4620} & \dualbest{0.4912}{\sigp}{0.0017} & \underline{0.5950} & 0.5821 & 0.5804 & \dualbest{0.6036}{\sigp}{0.0011} & 0.2297 & \underline{0.2325} & 0.2181 & \dualbest{0.2444}{\sigp}{0.0023} & 0.5006 & 0.4987 & \underline{0.5085} & \dualbest{0.5267}{\sigp}{0.0042} \\
\midrule
\multirow{4}{*}{GRU4Rec} & H@5 & 0.4717 & 0.4871 & \underline{0.5069} & \dualbest{0.5689}{\sigp}{0.0025} & 0.6942 & \underline{0.6943} & 0.6841 & \dualbest{0.7089}{\sigp}{0.0034} & 0.2938 & 0.2960 & \underline{0.2981} & \dualbest{0.3060}{\sigp}{0.0004} & 0.6188 & 0.6188 & \underline{0.6247} & \dualbest{0.6306}{\sigp}{0.0012} \\
 & H@10 & 0.5793 & 0.5819 & \underline{0.6078} & \dualbest{0.6836}{\sigp}{0.0019} & 0.7898 & \underline{0.7916} & 0.7824 & \dualbest{0.8015}{\sigp}{0.0021} & 0.4144 & \underline{0.4241} & 0.4232 & \dualbest{0.4310}{\sigp}{0.0019} & 0.7375 & 0.7418 & \underline{0.7524} & \dualbest{0.7687}{\sigp}{0.0013} \\
  & N@5 & 0.3632 & 0.3807 & \underline{0.3974} & \dualbest{0.4436}{\sigp}{0.0029} & 0.5529 & \underline{0.5543} & 0.5515 & \dualbest{0.5714}{\sigp}{0.0038} & 0.2006 & 0.2027 & \underline{0.2038} & \dualbest{0.2111}{\sigp}{0.0013} & \underline{0.4749} & 0.4730 & 0.4706 & \dualbest{0.4854}{\sigp}{0.0006} \\
  & N@10 & 0.3980 & 0.4114 & \underline{0.4301} & \dualbest{0.4808}{\sigp}{0.0023} & 0.5841 & \underline{0.5860} & 0.5825 & \dualbest{0.6015}{\sigp}{0.0027} & 0.2395 & 0.2441 & \underline{0.2452} & \dualbest{0.2513}{\sigp}{0.0017} & 0.5134 & \underline{0.5137} & 0.5123 & \dualbest{0.5254}{\sigp}{0.0007} \\
\bottomrule
\end{tabular}}
\end{table*}

\noindent\textbf{Search configuration.}
\method{} runs at most six outer rounds. Each round evaluates an initial
candidate at both rungs, followed by refinement beams of widths 3, 2, and 1.
Every beam is screened at low fidelity and its winner is evaluated at high
fidelity, giving seven candidate mechanisms per completed round.
Task-specific rung budgets and noise floors are fixed before search.
The high-fidelity budget is 300 epochs, while low-fidelity budgets range
from 15 to 295 epochs. The promotion-cost coefficient is $\lambda=0.5$.
Process weights $(w_D,w_E,w_A,w_N)=(0.35,0.15,0.30,0.20)$ and the
EMA coefficient $\alpha=0.3$ are shared across settings.
Appendix~\ref{app:setup} gives their values and calibration procedure.
The fixed stagnation settings invoke harness-mediated redirection after one
round without a candidate promotion. A second consecutive non-promoting round
stops the search. The four-round Fixed Rules and shortcut
studies specify their budgets in Appendix~\ref{app:additional-evidence}.

\subsection{RQ1: How Effective and Stable Is the Search?}
\label{sec:overall}

Table~\ref{tab:main_results} shows higher \method{} means than Base, OE, and
DE in all 64 setting--metric comparisons. Macro-averaging the relative
gains within settings gives improvements of 13.8\%, 10.0\%, 17.1\%, and
14.3\% over Base for H@5, H@10, N@5, and N@10.
The corresponding gains are 8.5\%, 6.1\%, 11.0\%, and 9.1\% over OE and
7.6\%, 6.8\%, 9.0\%, and 8.1\% over DE.
Improvements cover both hit rate and rank-sensitive accuracy across
the four recommendation methods.
On ML-1M/SASRec H@10, \method{} raises the strongest displayed baseline
mean from 0.8216 to 0.8239, corresponding to a 0.28\% relative gain.

\ScreeningEvidence
\begin{table}[!htbp]
\centering
\small
\caption{Representative changes in the archived final mechanisms.
Descriptions summarize executable changes recorded in the search traces.}
\label{tab:mechanisms}
\begin{tabularx}{\columnwidth}{@{}p{.31\columnwidth}>{\raggedright\arraybackslash}X@{}}
\toprule
Setting & Recorded change\\
\midrule
CDs\&Vinyl/NCF & Learnable frequency filtering of the history embedding sequence.\\
CDs\&Vinyl/NGCF & Simplified propagation with symmetric normalization.\\
CDs\&Vinyl/SASRec & Popularity correction of training-negative logits.\\
CDs\&Vinyl/GRU4Rec & Temperature annealing of the training logits.\\
Baby/NGCF & Co-annealed popularity debiasing and negative hardness.\\
Baby/SASRec & Stronger training-only sequence denoising.\\
Software/NCF & Two rates of exponential weight averaging.\\
Software/NGCF & Similarity-weighted graph propagation.\\
\bottomrule
\end{tabularx}
\end{table}

\begin{table}[!htbp]
\centering
\small
\setlength{\tabcolsep}{3pt}
\caption{Independent-search means over three searches. SD is shown for
\method{}; Table~\ref{tab:comparator_search_sd} supplies comparator SDs.
Base is the fixed mechanism in Table~\ref{tab:main_results}. Stars mark
unadjusted two-sided Welch significance against the higher-mean OE/DE
comparator: $^{*}p<0.05$, $^{**}p<0.01$, and $^{***}p<0.001$. Boldface and
underlining mark the best and second-best means.}
\label{tab:search_repeats}
\begin{tabular*}{\columnwidth}{@{\extracolsep{\fill}}lcccc@{}}
\toprule
Metric & Base & OE & DE & \method{} \\
\midrule
\multicolumn{5}{c}{Baby/NGCF (3 searches)} \\
\midrule
H@5 & 0.2131 & 0.2643 & \underline{0.2667} & $\mathbf{0.2798}\,\pm\,0.0048^{*}$ \\
H@10 & 0.3277 & \underline{0.3875} & 0.3844 & $\mathbf{0.3985}\,\pm\,0.0039^{*}$ \\
N@5 & 0.1387 & 0.1831 & \underline{0.1852} & $\mathbf{0.1994}\,\pm\,0.0058^{*}$ \\
N@10 & 0.1755 & 0.2212 & \underline{0.2228} & $\mathbf{0.2397}\,\pm\,0.0068^{*}$ \\
\midrule
\multicolumn{5}{c}{Software/NGCF (3 searches)} \\
\midrule
H@5 & 0.5356 & 0.5256 & \underline{0.5441} & $\mathbf{0.5598}\,\pm\,0.0058^{*}$ \\
H@10 & 0.6763 & 0.6604 & \underline{0.6797} & $\mathbf{0.6945}\,\pm\,0.0051^{*}$ \\
N@5 & 0.3918 & 0.3894 & \underline{0.4115} & $\mathbf{0.4278}\,\pm\,0.0025^{**}$ \\
N@10 & 0.4374 & 0.4331 & \underline{0.4508} & $\mathbf{0.4678}\,\pm\,0.0040^{**}$ \\
\midrule
\multicolumn{5}{c}{CDs\&Vinyl/GRU4Rec (3 searches)} \\
\midrule
H@5 & 0.4717 & 0.4872 & \underline{0.5068} & $\mathbf{0.5738}\,\pm\,0.0118^{**}$ \\
H@10 & 0.5793 & 0.5833 & \underline{0.6061} & $\mathbf{0.6903}\,\pm\,0.0101^{***}$ \\
N@5 & 0.3632 & 0.3787 & \underline{0.3972} & $\mathbf{0.4466}\,\pm\,0.0108^{**}$ \\
N@10 & 0.3980 & 0.4129 & \underline{0.4300} & $\mathbf{0.4844}\,\pm\,0.0102^{**}$ \\
\bottomrule
\end{tabular*}
\end{table}

\paragraph{Stability across independent searches.}
On three representative settings, we independently run OE, DE, and \method{}
three times with frozen GLM-5.2 under the shared 12-hour search cap.
OE and DE retain their native candidate schedules, while \method{} uses
at most six outer rounds. The runs share the three search-seed identifiers.
One search-level observation averages
the selected mechanism's test scores over training seeds 12352--12354.
Table~\ref{tab:search_repeats} reports the means and \method{} sample SDs
across these observations. Comparator SDs for the Welch tests appear in
Table~\ref{tab:comparator_search_sd}.

\method{} has the highest repeated-search mean in all 12 setting--metric
comparisons, with two-sided independent-sample Welch $p<0.05$ against the
higher-mean OE/DE comparator. Its between-search sample SDs span 0.0025--0.0118.

\subsection{RQ2: How Do Alternative Orchestration Designs Compare?}
\label{sec:ablation}

Full couples process-conditioned action-type utilities, posterior-sampled editing, and
evidence-conditioned harness redirection. We compare four executable
alternatives under the same six-round, 12-hour cap; Base and Full follow
Table~\ref{tab:main_results}.
\textbf{w/o Wiki} freezes the retained-candidate pool and consolidated
lesson store at their initial state. It stops accumulation of runner-ups
awaiting high-fidelity confirmation, budget-truncated promising candidates,
failure guidance, and outcome--diagnosis lessons. The incumbent and scored
trajectory remain available, and harness and skill rules can still evolve.

\textbf{w/o TS} retains the same actions, process rewards, and posterior
updates but replaces sampling with posterior-mean argmax.
\textbf{Fixed Rules} freezes rule writing: selected harness and skill edits
and evidence-conditioned redirection are disabled, while the two-round stop,
mechanism evolution, inside-wiki updates, process grading, and posterior
bookkeeping remain.

\textbf{Task-Gain} retains memory, Thompson sampling, editing, and the
stagnation response but replaces the process rubric with the within-round
change in the incumbent's validation objective. This task-gain signal enters
the same delayed update and action selection pipeline. All
comparisons retain the fixed task evaluator and mechanism admission rule.

\begin{table}[!tbp]
\centering
\small
\caption{Comparison of orchestration designs on two settings. Results are three-seed test
means. Full denotes \method{} in Table~\ref{tab:main_results}, and Base denotes
the unevolved mechanism. Boldface and underlining indicate the best and
second-best means for each metric within each setting.
\textit{w/o Wiki} freezes updates to the retained-candidate pool and
consolidated lessons.}
\label{tab:ablation}
\begin{tabular*}{\columnwidth}{@{\extracolsep{\fill}}lrrrr@{}}
\toprule
Design & H@5 & H@10 & N@5 & N@10 \\
\midrule
\multicolumn{5}{c}{Baby/NGCF} \\
\midrule
Full & \textbf{0.2784} & \textbf{0.3953} & \textbf{0.1896} & \textbf{0.2271} \\
w/o Wiki & \underline{0.2624} & \underline{0.3809} & \underline{0.1783} & \underline{0.2164} \\
w/o TS & 0.2574 & 0.3736 & 0.1757 & 0.2130 \\
Fixed Rules & 0.2250 & 0.3432 & 0.1480 & 0.1860 \\
Task-Gain & 0.2212 & 0.3310 & 0.1441 & 0.1804 \\
Base & 0.2131 & 0.3277 & 0.1387 & 0.1755 \\
\midrule
\multicolumn{5}{c}{CDs\&Vinyl/GRU4Rec} \\
\midrule
Full & \textbf{0.5689} & \textbf{0.6836} & \textbf{0.4436} & \textbf{0.4808} \\
w/o Wiki & \underline{0.5666} & 0.6703 & 0.4383 & 0.4718 \\
w/o TS & 0.5382 & 0.6424 & 0.4231 & 0.4569 \\
Fixed Rules & 0.5613 & \underline{0.6747} & 0.4352 & 0.4719 \\
Task-Gain & 0.5621 & 0.6732 & \underline{0.4392} & \underline{0.4752} \\
Base & 0.4717 & 0.5793 & 0.3632 & 0.3980 \\
\bottomrule
\end{tabular*}
\end{table}

\begin{figure}[t]
\centering
\includegraphics[trim=0bp 14bp 0bp 0bp,clip,width=\columnwidth]{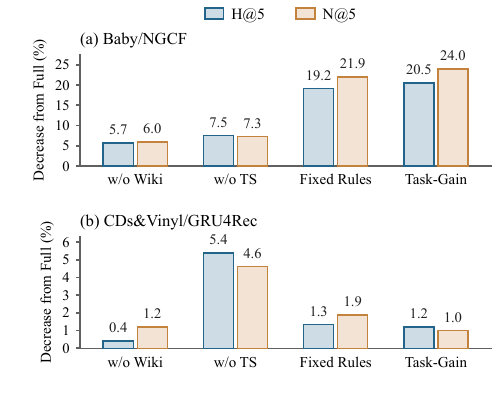}
\caption{Relative performance decrease from Full on H@5 and N@5.
Bars show $100(s_{\mathrm{Full}}-s_{\mathrm{variant}})/s_{\mathrm{Full}}$,
computed from the three-seed test means in Table~\ref{tab:ablation}.
The two panels use different y-axis ranges.}
\Description{Two vertically stacked panels compare the relative H@5 and N@5 decreases from
Full for w/o Wiki, w/o TS, Fixed Rules, and Task-Gain on Baby/NGCF and
CDs and Vinyl/GRU4Rec. Both vertical axes start at zero but use different
ranges.}
\label{fig:component_drop}
\end{figure}

Full achieves the highest score on all four metrics in both settings
(Table~\ref{tab:ablation}), with different gaps across variants.
Figure~\ref{fig:component_drop} visualizes the relative decreases in H@5
and N@5, the two metrics in the search objective.
On Baby/NGCF, Fixed Rules trails Full
by 0.0411--0.0534 across the four metrics. The losses are
0.0139--0.0217 for w/o TS and 0.0107--0.0160 for w/o Wiki.
On CDs\&Vinyl/GRU4Rec, w/o TS produces the largest gap
(0.0205--0.0412), compared with 0.0076--0.0089 for Fixed Rules
and 0.0023--0.0133 for w/o Wiki. Each alternative remains above Base.
The comparisons support structured memory, process-conditioned posterior
exploration over deterministic posterior-mean selection, and integrated
adaptive rule writing.

The reward comparison also differs across settings. Task-Gain trails
Full by 0.0455--0.0643 on Baby/NGCF, where its improvement over Base
is 0.0033--0.0081. On CDs\&Vinyl/GRU4Rec, its gap to Full is
0.0044--0.0104, and it remains 0.0760--0.0939 above Base.
With the same memory, delayed-update rule, selection policy, and edit
operators, process-rubric feedback yields higher test means than task-gain
feedback on both settings, with a larger gap on Baby/NGCF.

\ProcessValidationEvidence

\StressTestEvidence

\begin{figure}[t]
\centering
\includegraphics[trim=0bp 14bp 0bp 4bp,clip,width=\columnwidth]{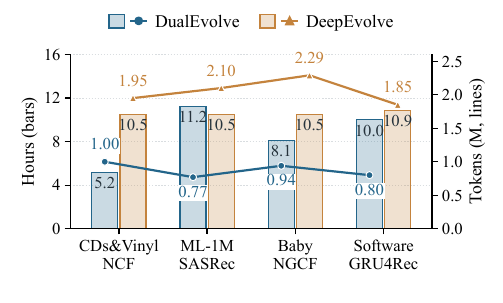}
\caption{Search cost on four settings. Bars show wall-clock hours (left
axis), and lines show LLM tokens in millions (right axis). DE token counts
are estimated.}
\Description{Grouped bars compare DualEvolve and DeepEvolve wall-clock hours,
while two lines compare their LLM token use on the right axis. The four
settings are CDs and Vinyl with NCF, MovieLens with SASRec, Baby with NGCF,
and Software with GRU4Rec.}
\label{fig:cost}
\end{figure}

\subsection{RQ3: Do Evolved Rules Transfer?}
\label{sec:transfer}

\begin{table}[!htbp]
\centering
\small
\setlength{\tabcolsep}{3pt}
\caption{Cross-setting transfer to Software/NGCF. Transferred rules and the
initial rules (No Transfer) are held fixed for one target-side round under the
same schedule. Values are three-seed test means; Base and Full come from
Table~\ref{tab:main_results}. Boldface and underlining mark the best and
second-best values. A star marks $p<0.05$ versus No Transfer by a two-sided
Welch $t$-test.}
\label{tab:transfer}
\begin{tabular*}{\columnwidth}{@{\extracolsep{\fill}}lcccc@{}}
\toprule
\textbf{Source / control} & \textbf{H@5} & \textbf{H@10} & \textbf{N@5} & \textbf{N@10} \\
\midrule
Baby/NGCF$^{\star}$ & 0.5403 & 0.6787 & 0.3932 & 0.4381 \\
Baby/NCF$^{\star}$ & \textbf{0.5605} & \textbf{0.6904} & \textbf{0.4198} & \textbf{0.4619} \\
No Transfer (1 round) & 0.5382 & 0.6775 & 0.3925 & 0.4378 \\
\quad Base & 0.5356 & 0.6763 & 0.3918 & 0.4374 \\
\quad Full & \underline{0.5530} & \underline{0.6876} & \underline{0.4132} & \underline{0.4569} \\
\bottomrule
\end{tabular*}
\end{table}

\begin{table}[!t]
\centering
\small
\caption{Source rules and corresponding target behavior. All transfers
target Software/NGCF.}
\label{tab:transfer_knowledge}
\begin{tabularx}{\columnwidth}{@{}p{.18\columnwidth}>{\raggedright\arraybackslash}X@{}}
\toprule
Source / layer & Rule and observed use\\
\midrule
Baby/NGCF\newline Harness &
\textbf{When progress is limited, favor less explored mechanism directions.}
The imported guidance initiates
optimizer search; subsequent target measurements guide a move to other
mechanisms.\\
\addlinespace[4.65pt]
Baby/NGCF\newline Skill &
\textbf{Continue refining an under-converged candidate.}
AdamW trails by 0.0015 at
140 epochs; its validation score rises from 0.5313 to 0.5367 at 300 epochs.
Later candidates adjust weight decay and training duration.\\
\addlinespace[4.65pt]
Baby/NCF\newline Harness &
\textbf{When progress is limited, favor less explored mechanism directions.}
After limited gains from DropEdge
refinements, the target tests losses and residual graph paths. A residual path raises the
validation objective from 0.5377 to 0.5558; tuning its weight reaches 0.5564.\\
\bottomrule
\end{tabularx}
\end{table}

We initialize Software/NGCF with source-evolved harness and skill rules
or the initial rules (No Transfer), holding them fixed for one target
round. Runs share the initial mechanism and start with a fresh inside
wiki. The transferred rules carry exploration summaries, while raw source
logs and wiki entries remain at the source. Each round uses Full's two-rung
schedule and candidate widths $1;3,2,1$. Table~\ref{tab:transfer} reports test
performance; source-rule acquisition costs are excluded.
Appendix~\ref{app:transfer_cost} gives additional transfer and cost details.

Both transfers significantly exceed No Transfer on all four metrics and
improve on target Base. Baby/NCF gains 2.1\%--7.1\% over Base and
0.4\%--1.6\% over target-specific Full; Baby/NGCF gains 0.2\%--0.9\%
over Base. Table~\ref{tab:transfer_knowledge} links transferred guidance to
target candidate sequences, including less explored directions and
continued refinement of under-converged candidates.

Target measurements determine revisions. In Baby/NGCF
transfer, the incumbent scores 0.5328 and AdamW 0.5313 at 140 epochs;
the latter reaches 0.5367 by epoch 300. The inside wiki records this
budget-specific evidence, while the transferred skill guides continued
refinement, including testing weight decay $10^{-4}$ instead of $10^{-2}$.
In Baby/NCF transfer, changing the residual weight from 0.1 to 0.75 raises
the validation objective from 0.5558 to 0.5564.

\subsection{RQ4: How Much Does the Search Cost?}
\label{sec:cost}

Figure~\ref{fig:cost} covers each dataset and recommendation method once.
\method{} uses 5.2--11.2 hours and 0.77--1.00 million tokens per search,
versus 10.5--10.9 hours and an estimated 1.85--2.29 million tokens for DE.
Token use is 49\%--63\% lower, and wall-clock time is lower in three settings.
ML-1M/\allowbreak SASRec takes 11.2 versus 10.5 hours, with a low-fidelity
budget of 295 out of 300 epochs. All frameworks share a 12-hour cap and
single-GPU configuration; actual runs can finish earlier.

\FloatBarrier
\section{Conclusion}
\label{sec:conclusion}

DualEvolve couples process-guided orchestration with noise- and cost-aware mechanism evolution under fixed evaluation and admission. It exceeds baseline test means in all 64 comparisons across 16 settings. Repeated searches and feedback comparisons assess the search design, while transfer results support reuse of evolved guidance. Token use is 49\%--63\% below estimated DeepEvolve on four settings.

\setlength{\bibsep}{0pt plus 1pt}
\bibliographystyle{ACM-Reference-Format}
\bibliography{references}


\begin{thebibliography}{45}


\ifx \showCODEN    \undefined \def \showCODEN     #1{\unskip}     \fi
\ifx \showISBNx    \undefined \def \showISBNx     #1{\unskip}     \fi
\ifx \showISBNxiii \undefined \def \showISBNxiii  #1{\unskip}     \fi
\ifx \showISSN     \undefined \def \showISSN      #1{\unskip}     \fi
\ifx \showLCCN     \undefined \def \showLCCN      #1{\unskip}     \fi
\ifx \shownote     \undefined \def \shownote      #1{#1}          \fi
\ifx \showarticletitle \undefined \def \showarticletitle #1{#1}   \fi
\ifx \showURL      \undefined \def \showURL       {\relax}        \fi
\providecommand\bibfield[2]{#2}
\providecommand\bibinfo[2]{#2}
\providecommand\natexlab[1]{#1}
\providecommand\showeprint[2][]{arXiv:#2}

\bibitem[Agrawal et~al\mbox{.}(2026)]%
        {agrawal2026gepa}
\bibfield{author}{\bibinfo{person}{Lakshya~A. Agrawal},
  \bibinfo{person}{Shangyin Tan}, \bibinfo{person}{Dilara Soylu},
  \bibinfo{person}{Noah Ziems}, \bibinfo{person}{Rishi Khare},
  \bibinfo{person}{Krista Opsahl{-}Ong}, \bibinfo{person}{Arnav Singhvi},
  \bibinfo{person}{Herumb Shandilya}, \bibinfo{person}{Michael~J. Ryan},
  \bibinfo{person}{Meng Jiang}, {et~al\mbox{.}}}
  \bibinfo{year}{2026}\natexlab{}.
\newblock \showarticletitle{{GEPA}: Reflective Prompt Evolution Can Outperform
  Reinforcement Learning}. In \bibinfo{booktitle}{\emph{The Fourteenth
  International Conference on Learning Representations}}.
\newblock
\urldef\tempurl%
\url{https://openreview.net/forum?id=RQm2KQTM5r}
\showURL{%
\tempurl}


\bibitem[Chan et~al\mbox{.}(2025)]%
        {DBLP:conf/iclr/ChanCJASMSLMPMW25}
\bibfield{author}{\bibinfo{person}{Jun~Shern Chan}, \bibinfo{person}{Neil
  Chowdhury}, \bibinfo{person}{Oliver Jaffe}, \bibinfo{person}{James Aung},
  \bibinfo{person}{Dane Sherburn}, \bibinfo{person}{Evan Mays},
  \bibinfo{person}{Giulio Starace}, \bibinfo{person}{Kevin Liu},
  \bibinfo{person}{Leon Maksin}, \bibinfo{person}{Tejal Patwardhan},
  {et~al\mbox{.}}} \bibinfo{year}{2025}\natexlab{}.
\newblock \showarticletitle{MLE-bench: Evaluating Machine Learning Agents on
  Machine Learning Engineering}. In \bibinfo{booktitle}{\emph{The Thirteenth
  International Conference on Learning Representations}}.
  \bibinfo{publisher}{OpenReview.net}.
\newblock
\urldef\tempurl%
\url{https://openreview.net/forum?id=6s5uXNWGIh}
\showURL{%
\tempurl}


\bibitem[Cheng et~al\mbox{.}(2024)]%
        {DBLP:conf/nips/ChengNS24}
\bibfield{author}{\bibinfo{person}{Ching{-}An Cheng}, \bibinfo{person}{Allen
  Nie}, {and} \bibinfo{person}{Adith Swaminathan}.}
  \bibinfo{year}{2024}\natexlab{}.
\newblock \showarticletitle{Trace is the Next {AutoDiff}: Generative
  Optimization with Rich Feedback, Execution Traces, and {LLM}s}. In
  \bibinfo{booktitle}{\emph{Advances in Neural Information Processing
  Systems}}, Vol.~\bibinfo{volume}{37}. \bibinfo{publisher}{Curran Associates,
  Inc.}, \bibinfo{pages}{71596--71642}.
\newblock
\href{https://doi.org/10.52202/079017-2287}{doi:\nolinkurl{10.52202/079017-2287}}


\bibitem[Dai et~al\mbox{.}(2026)]%
        {DBLP:journals/corr/abs-2606-24151}
\bibfield{author}{\bibinfo{person}{Zijie Dai}, \bibinfo{person}{Siuhin He},
  \bibinfo{person}{Hui Li}, \bibinfo{person}{Qihui Zhou},
  \bibinfo{person}{Jiajun Li}, \bibinfo{person}{Mingcong Song},
  \bibinfo{person}{Guoping Long}, \bibinfo{person}{Hongjie Si},
  \bibinfo{person}{Xin Yao}, \bibinfo{person}{Lin Zhang}, {et~al\mbox{.}}}
  \bibinfo{year}{2026}\natexlab{}.
\newblock \showarticletitle{Metis: Bridging Text and Code Memory for
  Self-Evolving Agents}.
\newblock \bibinfo{journal}{\emph{arXiv preprint arXiv:2606.24151}}
  (\bibinfo{year}{2026}).
\newblock


\bibitem[Guo et~al\mbox{.}(2024)]%
        {DBLP:conf/icml/GuoD0C0024}
\bibfield{author}{\bibinfo{person}{Siyuan Guo}, \bibinfo{person}{Cheng Deng},
  \bibinfo{person}{Ying Wen}, \bibinfo{person}{Hechang Chen},
  \bibinfo{person}{Yi Chang}, {and} \bibinfo{person}{Jun Wang}.}
  \bibinfo{year}{2024}\natexlab{}.
\newblock \showarticletitle{DS-Agent: Automated Data Science by Empowering
  Large Language Models with Case-Based Reasoning}. In
  \bibinfo{booktitle}{\emph{Proceedings of the 41st International Conference on
  Machine Learning}}, Vol.~\bibinfo{volume}{235}. \bibinfo{publisher}{{PMLR}},
  \bibinfo{pages}{16813--16848}.
\newblock
\urldef\tempurl%
\url{https://proceedings.mlr.press/v235/guo24b.html}
\showURL{%
\tempurl}


\bibitem[Hamadanian et~al\mbox{.}(2026)]%
        {DBLP:conf/cais/HamadanianKNNCP26}
\bibfield{author}{\bibinfo{person}{Pouya Hamadanian}, \bibinfo{person}{Pantea
  Karimi}, \bibinfo{person}{Arash Nasr{-}Esfahany}, \bibinfo{person}{Kimia
  Noorbakhsh}, \bibinfo{person}{Joseph Chandler}, \bibinfo{person}{Ali
  ParandehGheibi}, \bibinfo{person}{Mohammad Alizadeh}, {and}
  \bibinfo{person}{Hari Balakrishnan}.} \bibinfo{year}{2026}\natexlab{}.
\newblock \showarticletitle{Glia: A Human-Inspired {AI} for Automated Systems
  Design and Optimization}. In \bibinfo{booktitle}{\emph{Proceedings of the
  2026 {ACM} Conference on {AI} and Agentic Systems}}.
  \bibinfo{publisher}{{ACM}}, \bibinfo{pages}{61--84}.
\newblock
\href{https://doi.org/10.1145/3786335.3813125}{doi:\nolinkurl{10.1145/3786335.3813125}}


\bibitem[Han et~al\mbox{.}(2025)]%
        {DBLP:journals/corr/abs-2507-16075}
\bibfield{author}{\bibinfo{person}{Rujun Han}, \bibinfo{person}{Yanfei Chen},
  \bibinfo{person}{Zoey CuiZhu}, \bibinfo{person}{Lesly Miculicich},
  \bibinfo{person}{Guan Sun}, \bibinfo{person}{Yuanjun Bi},
  \bibinfo{person}{Weiming Wen}, \bibinfo{person}{Hui Wan},
  \bibinfo{person}{Chunfeng Wen}, \bibinfo{person}{Sol{\`e}ne Ma{\^i}tre},
  {et~al\mbox{.}}} \bibinfo{year}{2025}\natexlab{}.
\newblock \showarticletitle{Deep Researcher with Test-Time Diffusion}.
\newblock \bibinfo{journal}{\emph{arXiv preprint arXiv:2507.16075}}
  (\bibinfo{year}{2025}).
\newblock


\bibitem[Harper and Konstan(2016)]%
        {harper2015movielens}
\bibfield{author}{\bibinfo{person}{F.~Maxwell Harper} {and}
  \bibinfo{person}{Joseph~A. Konstan}.} \bibinfo{year}{2016}\natexlab{}.
\newblock \showarticletitle{The {MovieLens} Datasets: History and Context}.
\newblock \bibinfo{journal}{\emph{ACM Transactions on Interactive Intelligent
  Systems}} \bibinfo{volume}{5}, \bibinfo{number}{4} (\bibinfo{year}{2016}),
  \bibinfo{pages}{19:1--19:19}.
\newblock
\href{https://doi.org/10.1145/2827872}{doi:\nolinkurl{10.1145/2827872}}


\bibitem[He et~al\mbox{.}(2017)]%
        {he2017ncf}
\bibfield{author}{\bibinfo{person}{Xiangnan He}, \bibinfo{person}{Lizi Liao},
  \bibinfo{person}{Hanwang Zhang}, \bibinfo{person}{Liqiang Nie},
  \bibinfo{person}{Xia Hu}, {and} \bibinfo{person}{Tat{-}Seng Chua}.}
  \bibinfo{year}{2017}\natexlab{}.
\newblock \showarticletitle{Neural Collaborative Filtering}. In
  \bibinfo{booktitle}{\emph{Proceedings of the 26th International Conference on
  World Wide Web}}. \bibinfo{pages}{173--182}.
\newblock
\href{https://doi.org/10.1145/3038912.3052569}{doi:\nolinkurl{10.1145/3038912.3052569}}


\bibitem[Hidasi et~al\mbox{.}(2016)]%
        {hidasi2016gru4rec}
\bibfield{author}{\bibinfo{person}{Bal{\'a}zs Hidasi},
  \bibinfo{person}{Alexandros Karatzoglou}, \bibinfo{person}{Linas Baltrunas},
  {and} \bibinfo{person}{Domonkos Tikk}.} \bibinfo{year}{2016}\natexlab{}.
\newblock \showarticletitle{Session-Based Recommendations with Recurrent Neural
  Networks}. In \bibinfo{booktitle}{\emph{The Fourth International Conference
  on Learning Representations}}.
\newblock
\urldef\tempurl%
\url{https://arxiv.org/abs/1511.06939}
\showURL{%
\tempurl}


\bibitem[Hou et~al\mbox{.}(2026)]%
        {hou2024bridging}
\bibfield{author}{\bibinfo{person}{Yupeng Hou}, \bibinfo{person}{Jiacheng Li},
  \bibinfo{person}{Xiangjun Fu}, \bibinfo{person}{Zhankui He},
  \bibinfo{person}{An Yan}, \bibinfo{person}{Xiusi Chen}, {and}
  \bibinfo{person}{Julian McAuley}.} \bibinfo{year}{2026}\natexlab{}.
\newblock \showarticletitle{Bridging Language and Items for Retrieval and
  Recommendation: Benchmarking {LLM}s as Semantic Encoders}. In
  \bibinfo{booktitle}{\emph{Proceedings of the 64th Annual Meeting of the
  Association for Computational Linguistics (Volume 1: Long Papers)}}.
  \bibinfo{pages}{3251--3265}.
\newblock
\href{https://doi.org/10.18653/V1/2026.ACL-LONG.147}{doi:\nolinkurl{10.18653/V1/2026.ACL-LONG.147}}


\bibitem[Hu et~al\mbox{.}(2025)]%
        {DBLP:conf/iclr/HuLC25}
\bibfield{author}{\bibinfo{person}{Shengran Hu}, \bibinfo{person}{Cong Lu},
  {and} \bibinfo{person}{Jeff Clune}.} \bibinfo{year}{2025}\natexlab{}.
\newblock \showarticletitle{Automated Design of Agentic Systems}. In
  \bibinfo{booktitle}{\emph{The Thirteenth International Conference on Learning
  Representations}}. \bibinfo{publisher}{OpenReview.net}.
\newblock
\urldef\tempurl%
\url{https://openreview.net/forum?id=t9U3LW7JVX}
\showURL{%
\tempurl}


\bibitem[Huang et~al\mbox{.}(2024)]%
        {DBLP:conf/icml/HuangVLL24}
\bibfield{author}{\bibinfo{person}{Qian Huang}, \bibinfo{person}{Jian Vora},
  \bibinfo{person}{Percy Liang}, {and} \bibinfo{person}{Jure Leskovec}.}
  \bibinfo{year}{2024}\natexlab{}.
\newblock \showarticletitle{MLAgentBench: Evaluating Language Agents on Machine
  Learning Experimentation}. In \bibinfo{booktitle}{\emph{Proceedings of the
  41st International Conference on Machine Learning}},
  Vol.~\bibinfo{volume}{235}. \bibinfo{publisher}{{PMLR}},
  \bibinfo{pages}{20271--20309}.
\newblock
\urldef\tempurl%
\url{https://proceedings.mlr.press/v235/huang24y.html}
\showURL{%
\tempurl}


\bibitem[Kang and McAuley(2018)]%
        {kang2018sasrec}
\bibfield{author}{\bibinfo{person}{Wang{-}Cheng Kang} {and}
  \bibinfo{person}{Julian~J. McAuley}.} \bibinfo{year}{2018}\natexlab{}.
\newblock \showarticletitle{Self-Attentive Sequential Recommendation}. In
  \bibinfo{booktitle}{\emph{2018 IEEE International Conference on Data
  Mining}}. \bibinfo{pages}{197--206}.
\newblock
\href{https://doi.org/10.1109/ICDM.2018.00035}{doi:\nolinkurl{10.1109/ICDM.2018.00035}}


\bibitem[Kim et~al\mbox{.}(2026)]%
        {DBLP:journals/corr/abs-2602-12612}
\bibfield{author}{\bibinfo{person}{Sein Kim}, \bibinfo{person}{Sangwu Park},
  \bibinfo{person}{Hongseok Kang}, \bibinfo{person}{Wonjoong Kim},
  \bibinfo{person}{Jimin Seo}, \bibinfo{person}{Yeonjun In},
  \bibinfo{person}{Kanghoon Yoon}, \bibinfo{person}{Hyunsik Jeon}, {and}
  \bibinfo{person}{Chanyoung Park}.} \bibinfo{year}{2026}\natexlab{}.
\newblock \showarticletitle{Self-EvolveRec: Self-Evolving Recommender Systems
  with LLM-based Directional Feedback}.
\newblock \bibinfo{journal}{\emph{arXiv preprint arXiv:2602.12612}}
  (\bibinfo{year}{2026}).
\newblock


\bibitem[Lao et~al\mbox{.}(2026)]%
        {DBLP:journals/corr/abs-2606-26859}
\bibfield{author}{\bibinfo{person}{Changxin Lao}, \bibinfo{person}{Fei Pan},
  \bibinfo{person}{Guozhuang Ma}, \bibinfo{person}{Han Li},
  \bibinfo{person}{Huihuang Lin}, \bibinfo{person}{Jijun Shi},
  \bibinfo{person}{Kangzhi Zhao}, \bibinfo{person}{Kun Gai},
  \bibinfo{person}{Mo Zhou}, \bibinfo{person}{Qinqin Zhou}, {et~al\mbox{.}}}
  \bibinfo{year}{2026}\natexlab{}.
\newblock \showarticletitle{AgentX: Towards Agent-Driven Self-Iteration of
  Industrial Recommender Systems}.
\newblock \bibinfo{journal}{\emph{arXiv preprint arXiv:2606.26859}}
  (\bibinfo{year}{2026}).
\newblock


\bibitem[Lee et~al\mbox{.}(2026)]%
        {lee2026metaharness}
\bibfield{author}{\bibinfo{person}{Yoonho Lee}, \bibinfo{person}{Roshen Nair},
  \bibinfo{person}{Qizheng Zhang}, \bibinfo{person}{Kangwook Lee},
  \bibinfo{person}{Omar Khattab}, {and} \bibinfo{person}{Chelsea Finn}.}
  \bibinfo{year}{2026}\natexlab{}.
\newblock \showarticletitle{{Meta-Harness}: End-to-End Optimization of Model
  Harnesses}.
\newblock \bibinfo{journal}{\emph{arXiv preprint arXiv:2603.28052}}
  (\bibinfo{year}{2026}).
\newblock


\bibitem[Li et~al\mbox{.}(2025)]%
        {DBLP:conf/nips/LiSYZ25}
\bibfield{author}{\bibinfo{person}{Zhening Li}, \bibinfo{person}{Armando
  Solar{-}Lezama}, \bibinfo{person}{Yisong Yue}, {and} \bibinfo{person}{Stephan
  Zheng}.} \bibinfo{year}{2025}\natexlab{}.
\newblock \showarticletitle{{EnCompass}: Enhancing Agent Programming with
  Search Over Program Execution Paths}. In \bibinfo{booktitle}{\emph{Advances
  in Neural Information Processing Systems}}, Vol.~\bibinfo{volume}{38}.
  \bibinfo{publisher}{Curran Associates, Inc.},
  \bibinfo{pages}{105751--105815}.
\newblock
\href{https://doi.org/10.52202/085713-3533}{doi:\nolinkurl{10.52202/085713-3533}}


\bibitem[Lin et~al\mbox{.}(2026)]%
        {lin2026agenticharness}
\bibfield{author}{\bibinfo{person}{Jiahang Lin}, \bibinfo{person}{Shichun Liu},
  \bibinfo{person}{Chengjun Pan}, \bibinfo{person}{Lizhi Lin},
  \bibinfo{person}{Shihan Dou}, \bibinfo{person}{Zhiheng Xi},
  \bibinfo{person}{Xuanjing Huang}, \bibinfo{person}{Hang Yan},
  \bibinfo{person}{Zhenhua Han}, \bibinfo{person}{Tao Gui}, {et~al\mbox{.}}}
  \bibinfo{year}{2026}\natexlab{}.
\newblock \showarticletitle{Agentic Harness Engineering: Observability-Driven
  Automatic Evolution of Coding-Agent Harnesses}.
\newblock \bibinfo{journal}{\emph{arXiv preprint arXiv:2604.25850}}
  (\bibinfo{year}{2026}).
\newblock


\bibitem[Liu et~al\mbox{.}(2020)]%
        {DBLP:conf/kdd/LiuZL0LTHL020}
\bibfield{author}{\bibinfo{person}{Bin Liu}, \bibinfo{person}{Chenxu Zhu},
  \bibinfo{person}{Guilin Li}, \bibinfo{person}{Weinan Zhang},
  \bibinfo{person}{Jincai Lai}, \bibinfo{person}{Ruiming Tang},
  \bibinfo{person}{Xiuqiang He}, \bibinfo{person}{Zhenguo Li}, {and}
  \bibinfo{person}{Yong Yu}.} \bibinfo{year}{2020}\natexlab{}.
\newblock \showarticletitle{{AutoFIS}: Automatic Feature Interaction Selection
  in Factorization Models for Click-Through Rate Prediction}. In
  \bibinfo{booktitle}{\emph{Proceedings of the 26th {ACM} {SIGKDD}
  International Conference on Knowledge Discovery and Data Mining}}.
  \bibinfo{publisher}{{ACM}}, \bibinfo{pages}{2636--2645}.
\newblock
\href{https://doi.org/10.1145/3394486.3403314}{doi:\nolinkurl{10.1145/3394486.3403314}}


\bibitem[Liu et~al\mbox{.}(2025)]%
        {DBLP:journals/corr/abs-2510-06056}
\bibfield{author}{\bibinfo{person}{Gang Liu}, \bibinfo{person}{Yihan Zhu},
  \bibinfo{person}{Jie Chen}, {and} \bibinfo{person}{Meng Jiang}.}
  \bibinfo{year}{2025}\natexlab{}.
\newblock \showarticletitle{Scientific Algorithm Discovery by Augmenting
  AlphaEvolve with Deep Research}.
\newblock \bibinfo{journal}{\emph{arXiv preprint arXiv:2510.06056}}
  (\bibinfo{year}{2025}).
\newblock


\bibitem[Liu et~al\mbox{.}(2026)]%
        {DBLP:journals/corr/abs-2606-27243}
\bibfield{author}{\bibinfo{person}{Shaohua Liu}, \bibinfo{person}{Liang Fang},
  \bibinfo{person}{Yilong Sun}, \bibinfo{person}{Shudong Huang},
  \bibinfo{person}{Qingsong Luo}, \bibinfo{person}{Shaoxin Liu},
  \bibinfo{person}{Xiaoyang Chen}, \bibinfo{person}{Dongqiang Liu},
  \bibinfo{person}{Chuangang Ma}, \bibinfo{person}{Zhenzhen Chai},
  {et~al\mbox{.}}} \bibinfo{year}{2026}\natexlab{}.
\newblock \showarticletitle{{NOVA:} {A} Verification-Aware Agent Harness for
  Architecture Evolution in Industrial Recommender Systems}.
\newblock \bibinfo{journal}{\emph{arXiv preprint arXiv:2606.27243}}
  (\bibinfo{year}{2026}).
\newblock


\bibitem[Lu et~al\mbox{.}(2026)]%
        {DBLP:journals/corr/abs-2606-04455}
\bibfield{author}{\bibinfo{person}{Xinyu Lu}, \bibinfo{person}{Tianshu Wang},
  \bibinfo{person}{Pengbo Wang}, \bibinfo{person}{Zujie Wen},
  \bibinfo{person}{Zhiqiang Zhang}, \bibinfo{person}{Jun Zhou},
  \bibinfo{person}{Boxi Cao}, \bibinfo{person}{Yaojie Lu},
  \bibinfo{person}{Hongyu Lin}, \bibinfo{person}{Xianpei Han}, {et~al\mbox{.}}}
  \bibinfo{year}{2026}\natexlab{}.
\newblock \showarticletitle{The Meta-Agent Challenge: Are Current Agents
  Capable of Autonomous Agent Development?}
\newblock \bibinfo{journal}{\emph{arXiv preprint arXiv:2606.04455}}
  (\bibinfo{year}{2026}).
\newblock


\bibitem[Luo et~al\mbox{.}(2025)]%
        {DBLP:journals/corr/abs-2508-03680}
\bibfield{author}{\bibinfo{person}{Xufang Luo}, \bibinfo{person}{Yuge Zhang},
  \bibinfo{person}{Zhiyuan He}, \bibinfo{person}{Zilong Wang},
  \bibinfo{person}{Siyun Zhao}, \bibinfo{person}{Dongsheng Li},
  \bibinfo{person}{Luna~K. Qiu}, {and} \bibinfo{person}{Yuqing Yang}.}
  \bibinfo{year}{2025}\natexlab{}.
\newblock \showarticletitle{Agent Lightning: Train {ANY} {AI} Agents with
  Reinforcement Learning}.
\newblock \bibinfo{journal}{\emph{arXiv preprint arXiv:2508.03680}}
  (\bibinfo{year}{2025}).
\newblock


\bibitem[Madaan et~al\mbox{.}(2023)]%
        {DBLP:conf/nips/MadaanTGHGW0DPY23}
\bibfield{author}{\bibinfo{person}{Aman Madaan}, \bibinfo{person}{Niket
  Tandon}, \bibinfo{person}{Prakhar Gupta}, \bibinfo{person}{Skyler Hallinan},
  \bibinfo{person}{Luyu Gao}, \bibinfo{person}{Sarah Wiegreffe},
  \bibinfo{person}{Uri Alon}, \bibinfo{person}{Nouha Dziri},
  \bibinfo{person}{Shrimai Prabhumoye}, \bibinfo{person}{Yiming Yang},
  {et~al\mbox{.}}} \bibinfo{year}{2023}\natexlab{}.
\newblock \showarticletitle{Self-Refine: Iterative Refinement with
  Self-Feedback}. In \bibinfo{booktitle}{\emph{Advances in Neural Information
  Processing Systems}}, Vol.~\bibinfo{volume}{36}. \bibinfo{publisher}{Curran
  Associates, Inc.}, \bibinfo{pages}{46534--46594}.
\newblock
\href{https://doi.org/10.52202/075280-2019}{doi:\nolinkurl{10.52202/075280-2019}}


\bibitem[Nam et~al\mbox{.}(2025)]%
        {DBLP:conf/nips/NamYCSAP25}
\bibfield{author}{\bibinfo{person}{Jaehyun Nam}, \bibinfo{person}{Jinsung
  Yoon}, \bibinfo{person}{Jiefeng Chen}, \bibinfo{person}{Jinwoo Shin},
  \bibinfo{person}{Sercan~{\"{O}}. Arik}, {and} \bibinfo{person}{Tomas
  Pfister}.} \bibinfo{year}{2025}\natexlab{}.
\newblock \showarticletitle{{MLE-STAR}: Machine Learning Engineering Agent via
  Search and Targeted Refinement}. In \bibinfo{booktitle}{\emph{Advances in
  Neural Information Processing Systems}}, Vol.~\bibinfo{volume}{38}.
  \bibinfo{publisher}{Curran Associates, Inc.},
  \bibinfo{pages}{116692--116712}.
\newblock
\href{https://doi.org/10.52202/085713-3889}{doi:\nolinkurl{10.52202/085713-3889}}


\bibitem[Nie et~al\mbox{.}(2026)]%
        {DBLP:journals/corr/abs-2607-08124}
\bibfield{author}{\bibinfo{person}{Jun Nie}, \bibinfo{person}{Yonggang Zhang},
  \bibinfo{person}{Jun Song}, \bibinfo{person}{Qianshu Cai},
  \bibinfo{person}{Dahai Yu}, \bibinfo{person}{Yike Guo},
  \bibinfo{person}{Xinmei Tian}, {and} \bibinfo{person}{Bo Han}.}
  \bibinfo{year}{2026}\natexlab{}.
\newblock \showarticletitle{TTHE: Test-Time Harness Evolution}.
\newblock \bibinfo{journal}{\emph{arXiv preprint arXiv:2607.08124}}
  (\bibinfo{year}{2026}).
\newblock


\bibitem[Novikov et~al\mbox{.}(2025)]%
        {DBLP:journals/corr/abs-2506-13131}
\bibfield{author}{\bibinfo{person}{Alexander Novikov},
  \bibinfo{person}{Ng{\^a}n V{\~u}}, \bibinfo{person}{Marvin Eisenberger},
  \bibinfo{person}{Emilien Dupont}, \bibinfo{person}{Po{-}Sen Huang},
  \bibinfo{person}{Adam~Zsolt Wagner}, \bibinfo{person}{Sergey Shirobokov},
  \bibinfo{person}{Borislav Kozlovskii}, \bibinfo{person}{Francisco J.~R.
  Ruiz}, \bibinfo{person}{Abbas Mehrabian}, {et~al\mbox{.}}}
  \bibinfo{year}{2025}\natexlab{}.
\newblock \showarticletitle{AlphaEvolve: {A} coding agent for scientific and
  algorithmic discovery}.
\newblock \bibinfo{journal}{\emph{arXiv preprint arXiv:2506.13131}}
  (\bibinfo{year}{2025}).
\newblock


\bibitem[Romera{-}Paredes et~al\mbox{.}(2024)]%
        {romeraparedes2024funsearch}
\bibfield{author}{\bibinfo{person}{Bernardino Romera{-}Paredes},
  \bibinfo{person}{Mohammadamin Barekatain}, \bibinfo{person}{Alexander
  Novikov}, \bibinfo{person}{Matej Balog}, \bibinfo{person}{M.~Pawan Kumar},
  \bibinfo{person}{Emilien Dupont}, \bibinfo{person}{Francisco J.~R. Ruiz},
  \bibinfo{person}{Jordan~S. Ellenberg}, \bibinfo{person}{Pengming Wang},
  \bibinfo{person}{Omar Fawzi}, {et~al\mbox{.}}}
  \bibinfo{year}{2024}\natexlab{}.
\newblock \showarticletitle{Mathematical discoveries from program search with
  large language models}.
\newblock \bibinfo{journal}{\emph{Nature}} \bibinfo{volume}{625},
  \bibinfo{number}{7995} (\bibinfo{year}{2024}), \bibinfo{pages}{468--475}.
\newblock
\href{https://doi.org/10.1038/s41586-023-06924-6}{doi:\nolinkurl{10.1038/s41586-023-06924-6}}


\bibitem[Shang et~al\mbox{.}(2025)]%
        {DBLP:conf/iclr/ShangLZMLXL25}
\bibfield{author}{\bibinfo{person}{Yu Shang}, \bibinfo{person}{Yu Li},
  \bibinfo{person}{Keyu Zhao}, \bibinfo{person}{Likai Ma},
  \bibinfo{person}{Jiahe Liu}, \bibinfo{person}{Fengli Xu}, {and}
  \bibinfo{person}{Yong Li}.} \bibinfo{year}{2025}\natexlab{}.
\newblock \showarticletitle{AgentSquare: Automatic {LLM} Agent Search in
  Modular Design Space}. In \bibinfo{booktitle}{\emph{The Thirteenth
  International Conference on Learning Representations}}.
  \bibinfo{publisher}{OpenReview.net}.
\newblock
\urldef\tempurl%
\url{https://openreview.net/forum?id=mPdmDYIQ7f}
\showURL{%
\tempurl}


\bibitem[Shao et~al\mbox{.}(2026)]%
        {DBLP:journals/corr/abs-2509-26354}
\bibfield{author}{\bibinfo{person}{Shuai Shao}, \bibinfo{person}{Qihan Ren},
  \bibinfo{person}{Dongrui Liu}, \bibinfo{person}{Chen Qian},
  \bibinfo{person}{Boyi Wei}, \bibinfo{person}{Dadi Guo},
  \bibinfo{person}{Jingyi Yang}, \bibinfo{person}{Xinhao Song},
  \bibinfo{person}{Linfeng Zhang}, \bibinfo{person}{Weinan Zhang},
  {et~al\mbox{.}}} \bibinfo{year}{2026}\natexlab{}.
\newblock \showarticletitle{Your Agent May Misevolve: Emergent Risks in
  Self-evolving {LLM} Agents}. In \bibinfo{booktitle}{\emph{The Fourteenth
  International Conference on Learning Representations}}.
  \bibinfo{pages}{99728--99793}.
\newblock
\urldef\tempurl%
\url{https://proceedings.iclr.cc/paper_files/paper/2026/hash/a24cd16bc361afa78e57d31d34f3d936-Abstract-Conference.html}
\showURL{%
\tempurl}


\bibitem[Sharma(2025)]%
        {sharma2025openevolve}
\bibfield{author}{\bibinfo{person}{Asankhaya Sharma}.}
  \bibinfo{year}{2025}\natexlab{}.
\newblock \bibinfo{title}{OpenEvolve: an open-source evolutionary coding
  agent}.
\newblock
  \bibinfo{howpublished}{\url{https://github.com/algorithmicsuperintelligence/openevolve}}.
\newblock
\shownote{GitHub repository}.
\newblock


\bibitem[Shinn et~al\mbox{.}(2023)]%
        {DBLP:conf/nips/ShinnCGNY23}
\bibfield{author}{\bibinfo{person}{Noah Shinn}, \bibinfo{person}{Federico
  Cassano}, \bibinfo{person}{Ashwin Gopinath}, \bibinfo{person}{Karthik
  Narasimhan}, {and} \bibinfo{person}{Shunyu Yao}.}
  \bibinfo{year}{2023}\natexlab{}.
\newblock \showarticletitle{Reflexion: Language Agents with Verbal
  Reinforcement Learning}. In \bibinfo{booktitle}{\emph{Advances in Neural
  Information Processing Systems}}, Vol.~\bibinfo{volume}{36}.
  \bibinfo{publisher}{Curran Associates, Inc.}, \bibinfo{pages}{8634--8652}.
\newblock
\href{https://doi.org/10.52202/075280-0377}{doi:\nolinkurl{10.52202/075280-0377}}


\bibitem[Song et~al\mbox{.}(2020)]%
        {DBLP:conf/kdd/SongCZYTH20}
\bibfield{author}{\bibinfo{person}{Qingquan Song}, \bibinfo{person}{Dehua
  Cheng}, \bibinfo{person}{Hanning Zhou}, \bibinfo{person}{Jiyan Yang},
  \bibinfo{person}{Yuandong Tian}, {and} \bibinfo{person}{Xia Hu}.}
  \bibinfo{year}{2020}\natexlab{}.
\newblock \showarticletitle{Towards Automated Neural Interaction Discovery for
  Click-Through Rate Prediction}. In \bibinfo{booktitle}{\emph{Proceedings of
  the 26th {ACM} {SIGKDD} International Conference on Knowledge Discovery and
  Data Mining}}. \bibinfo{publisher}{{ACM}}, \bibinfo{pages}{945--955}.
\newblock
\href{https://doi.org/10.1145/3394486.3403137}{doi:\nolinkurl{10.1145/3394486.3403137}}


\bibitem[Starace et~al\mbox{.}(2025)]%
        {DBLP:conf/icml/StaraceJSACMDMK25}
\bibfield{author}{\bibinfo{person}{Giulio Starace}, \bibinfo{person}{Oliver
  Jaffe}, \bibinfo{person}{Dane Sherburn}, \bibinfo{person}{James Aung},
  \bibinfo{person}{Jun~Shern Chan}, \bibinfo{person}{Leon Maksin},
  \bibinfo{person}{Rachel Dias}, \bibinfo{person}{Evan Mays},
  \bibinfo{person}{Benjamin Kinsella}, \bibinfo{person}{Wyatt Thompson},
  {et~al\mbox{.}}} \bibinfo{year}{2025}\natexlab{}.
\newblock \showarticletitle{PaperBench: Evaluating AI's Ability to Replicate
  {AI} Research}. In \bibinfo{booktitle}{\emph{Proceedings of the 42nd
  International Conference on Machine Learning}}, Vol.~\bibinfo{volume}{267}.
  \bibinfo{publisher}{{PMLR}}, \bibinfo{pages}{56843--56873}.
\newblock
\urldef\tempurl%
\url{https://proceedings.mlr.press/v267/starace25a.html}
\showURL{%
\tempurl}


\bibitem[Tao et~al\mbox{.}(2026)]%
        {DBLP:journals/corr/abs-2606-04465}
\bibfield{author}{\bibinfo{person}{Wangcheng Tao}, \bibinfo{person}{Han Wu},
  {and} \bibinfo{person}{Weng{-}Fai Wong}.} \bibinfo{year}{2026}\natexlab{}.
\newblock \showarticletitle{SePO: Self-Evolving Prompt Agent for System Prompt
  Optimization}.
\newblock \bibinfo{journal}{\emph{arXiv preprint arXiv:2606.04465}}
  (\bibinfo{year}{2026}).
\newblock


\bibitem[Wang et~al\mbox{.}(2019)]%
        {wang2019ngcf}
\bibfield{author}{\bibinfo{person}{Xiang Wang}, \bibinfo{person}{Xiangnan He},
  \bibinfo{person}{Meng Wang}, \bibinfo{person}{Fuli Feng}, {and}
  \bibinfo{person}{Tat{-}Seng Chua}.} \bibinfo{year}{2019}\natexlab{}.
\newblock \showarticletitle{Neural Graph Collaborative Filtering}. In
  \bibinfo{booktitle}{\emph{Proceedings of the 42nd International ACM SIGIR
  Conference on Research and Development in Information Retrieval}}.
  \bibinfo{pages}{165--174}.
\newblock
\href{https://doi.org/10.1145/3331184.3331267}{doi:\nolinkurl{10.1145/3331184.3331267}}


\bibitem[Xia et~al\mbox{.}(2025)]%
        {DBLP:journals/corr/abs-2511-13646}
\bibfield{author}{\bibinfo{person}{Chunqiu~Steven Xia}, \bibinfo{person}{Zhe
  Wang}, \bibinfo{person}{Yan Yang}, \bibinfo{person}{Yuxiang Wei}, {and}
  \bibinfo{person}{Lingming Zhang}.} \bibinfo{year}{2025}\natexlab{}.
\newblock \showarticletitle{Live-SWE-agent: Can Software Engineering Agents
  Self-Evolve on the Fly?}
\newblock \bibinfo{journal}{\emph{arXiv preprint arXiv:2511.13646}}
  (\bibinfo{year}{2025}).
\newblock


\bibitem[Yin et~al\mbox{.}(2025)]%
        {DBLP:conf/acl/YinWPL0W25}
\bibfield{author}{\bibinfo{person}{Xunjian Yin}, \bibinfo{person}{Xinyi Wang},
  \bibinfo{person}{Liangming Pan}, \bibinfo{person}{Li Lin},
  \bibinfo{person}{Xiaojun Wan}, {and} \bibinfo{person}{William~Yang Wang}.}
  \bibinfo{year}{2025}\natexlab{}.
\newblock \showarticletitle{G{\"{o}}del Agent: {A} Self-Referential Agent
  Framework for Recursively Self-Improvement}. In
  \bibinfo{booktitle}{\emph{Proceedings of the 63rd Annual Meeting of the
  Association for Computational Linguistics (Volume 1: Long Papers), {ACL}
  2025, Vienna, Austria, July 27 - August 1, 2025}}.
  \bibinfo{publisher}{Association for Computational Linguistics},
  \bibinfo{pages}{27890--27913}.
\newblock
\href{https://doi.org/10.18653/V1/2025.ACL-LONG.1354}{doi:\nolinkurl{10.18653/V1/2025.ACL-LONG.1354}}


\bibitem[Zhang et~al\mbox{.}(2024a)]%
        {DBLP:conf/sigir/0003CSWC24}
\bibfield{author}{\bibinfo{person}{An Zhang}, \bibinfo{person}{Yuxin Chen},
  \bibinfo{person}{Leheng Sheng}, \bibinfo{person}{Xiang Wang}, {and}
  \bibinfo{person}{Tat{-}Seng Chua}.} \bibinfo{year}{2024}\natexlab{a}.
\newblock \showarticletitle{On Generative Agents in Recommendation}. In
  \bibinfo{booktitle}{\emph{Proceedings of the 47th International {ACM} {SIGIR}
  Conference on Research and Development in Information Retrieval}}.
  \bibinfo{publisher}{{ACM}}, \bibinfo{pages}{1807--1817}.
\newblock
\href{https://doi.org/10.1145/3626772.3657844}{doi:\nolinkurl{10.1145/3626772.3657844}}


\bibitem[Zhang et~al\mbox{.}(2025a)]%
        {DBLP:conf/icml/ZhangNF00025}
\bibfield{author}{\bibinfo{person}{Guibin Zhang}, \bibinfo{person}{Luyang Niu},
  \bibinfo{person}{Junfeng Fang}, \bibinfo{person}{Kun Wang},
  \bibinfo{person}{Lei Bai}, {and} \bibinfo{person}{Xiang Wang}.}
  \bibinfo{year}{2025}\natexlab{a}.
\newblock \showarticletitle{Multi-agent Architecture Search via Agentic
  Supernet}. In \bibinfo{booktitle}{\emph{Proceedings of the 42nd International
  Conference on Machine Learning}}, Vol.~\bibinfo{volume}{267}.
  \bibinfo{publisher}{{PMLR}}, \bibinfo{pages}{75834--75852}.
\newblock
\urldef\tempurl%
\url{https://proceedings.mlr.press/v267/zhang25bi.html}
\showURL{%
\tempurl}


\bibitem[Zhang et~al\mbox{.}(2024b)]%
        {DBLP:conf/www/ZhangHXSMZLW24}
\bibfield{author}{\bibinfo{person}{Junjie Zhang}, \bibinfo{person}{Yupeng Hou},
  \bibinfo{person}{Ruobing Xie}, \bibinfo{person}{Wenqi Sun},
  \bibinfo{person}{Julian~J. McAuley}, \bibinfo{person}{Wayne~Xin Zhao},
  \bibinfo{person}{Leyu Lin}, {and} \bibinfo{person}{Ji{-}Rong Wen}.}
  \bibinfo{year}{2024}\natexlab{b}.
\newblock \showarticletitle{{AgentCF}: Collaborative Learning with Autonomous
  Language Agents for Recommender Systems}. In
  \bibinfo{booktitle}{\emph{Proceedings of the {ACM} Web Conference 2024}}.
  \bibinfo{publisher}{{ACM}}, \bibinfo{pages}{3679--3689}.
\newblock
\href{https://doi.org/10.1145/3589334.3645537}{doi:\nolinkurl{10.1145/3589334.3645537}}


\bibitem[Zhang et~al\mbox{.}(2026b)]%
        {DBLP:journals/corr/abs-2505-22954}
\bibfield{author}{\bibinfo{person}{Jenny Zhang}, \bibinfo{person}{Shengran Hu},
  \bibinfo{person}{Cong Lu}, \bibinfo{person}{Robert Lange}, {and}
  \bibinfo{person}{Jeff Clune}.} \bibinfo{year}{2026}\natexlab{b}.
\newblock \showarticletitle{Darwin G{\"o}del Machine: Open-Ended Evolution of
  Self-Improving Agents}. In \bibinfo{booktitle}{\emph{The Fourteenth
  International Conference on Learning Representations}}.
  \bibinfo{pages}{104223--104294}.
\newblock
\urldef\tempurl%
\url{https://proceedings.iclr.cc/paper_files/paper/2026/hash/aa5f5e6eb6f613ec412f1d948dfa21a5-Abstract-Conference.html}
\showURL{%
\tempurl}


\bibitem[Zhang et~al\mbox{.}(2025b)]%
        {DBLP:conf/iclr/ZhangXYTCCZCHWZ25}
\bibfield{author}{\bibinfo{person}{Jiayi Zhang}, \bibinfo{person}{Jinyu Xiang},
  \bibinfo{person}{Zhaoyang Yu}, \bibinfo{person}{Fengwei Teng},
  \bibinfo{person}{Xiong{-}Hui Chen}, \bibinfo{person}{Jiaqi Chen},
  \bibinfo{person}{Mingchen Zhuge}, \bibinfo{person}{Xin Cheng},
  \bibinfo{person}{Sirui Hong}, \bibinfo{person}{Jinlin Wang}, {et~al\mbox{.}}}
  \bibinfo{year}{2025}\natexlab{b}.
\newblock \showarticletitle{AFlow: Automating Agentic Workflow Generation}. In
  \bibinfo{booktitle}{\emph{The Thirteenth International Conference on Learning
  Representations}}. \bibinfo{publisher}{OpenReview.net}.
\newblock
\urldef\tempurl%
\url{https://openreview.net/forum?id=z5uVAKwmjf}
\showURL{%
\tempurl}


\bibitem[Zhang et~al\mbox{.}(2026a)]%
        {DBLP:journals/corr/abs-2605-22148}
\bibfield{author}{\bibinfo{person}{Xing Zhang}, \bibinfo{person}{Yanwei Cui},
  \bibinfo{person}{Guanghui Wang}, \bibinfo{person}{Ziyuan Li},
  \bibinfo{person}{Wei Qiu}, \bibinfo{person}{Bing Zhu}, {and}
  \bibinfo{person}{Peiyang He}.} \bibinfo{year}{2026}\natexlab{a}.
\newblock \showarticletitle{Ratchet: How Reliable Must an {LLM} Judge Be to
  Retire a Skill?}
\newblock \bibinfo{journal}{\emph{arXiv preprint arXiv:2605.22148}}
  (\bibinfo{year}{2026}).
\newblock


\end{thebibliography}

\appendix
\section{Experimental Details}
\label{app:setup}

\begin{table}[t]
\centering
\caption{Dataset statistics after preprocessing. \#Train denotes the number
of training interactions; \#Valid and \#Test denote the numbers of users
whose held-out items remain after filtering. Average sequence length is
computed over all retained interactions.}
\label{tab:datasets}
\begin{tabular*}{\columnwidth}{@{\extracolsep{\fill}}lrrrr@{}}
\toprule
 & \textbf{CDs\&Vinyl} & \textbf{ML-1M} & \textbf{Baby} & \textbf{Software} \\
\midrule
\#Users        & 14{,}335 & 6{,}040 & 14{,}335 & 14{,}335 \\
\#Items        & 11{,}436 & 3{,}416 & 6{,}228 & 3{,}816 \\
\#Train        & 106{,}449 & 987{,}538 & 85{,}103 & 91{,}722 \\
\#Valid        & 10{,}060 & 6{,}037 & 11{,}974 & 13{,}255 \\
\#Test         & 9{,}716 & 6{,}036 & 11{,}564 & 13{,}250 \\
Avg.\ seq.\ len & 8.81 & 165.50 & 7.58 & 8.25 \\
\bottomrule
\end{tabular*}
\end{table}

\paragraph{Fixed settings.}
Table~\ref{tab:datasets} gives post-filtering statistics. Rung budgets and
task-specific thresholds are fixed before search and remain outside
the editable orchestration. All recommendation searches use a frozen
GLM-5.2. Table~\ref{tab:global_parameters} lists shared control parameters;
Table~\ref{tab:task_calibration} provides the task configurations for reproduction.

\begin{table}[!ht]
\centering
\footnotesize
\caption{Global search parameters. Each process dimension averages three
returned grades.}
\label{tab:global_parameters}
\begin{tabular*}{\columnwidth}{@{\extracolsep{\fill}}lr@{}}
\toprule
Parameter & Value\\
\midrule
Maximum outer rounds & 6\\
Non-promoting rounds to redirect / stop & 1 / 2\\
Initial candidate and refinement beam widths & $1;\ 3,2,1$\\
Backtracking threshold $K$ & 1\\
Promotion-cost coefficient $\lambda$ & 0.5\\
Process weights $(w_D,w_E,w_A,w_N)$ & $(0.35,0.15,0.30,0.20)$\\
Grading repetitions per dimension & 3\\
EMA coefficient $\alpha$ & 0.3\\
No-measurement fraction / measured-count cap & 0.5 / 2\\
Failure-penalty coefficient & 0.5\\
Prior variance $\tau_0^2$ & 1.0\\
Observation / sampling variance floors & $10^{-4}$ / 0.05\\
Main-table test seeds & 12352, 12353, 12354\\
\bottomrule
\end{tabular*}
\end{table}

\begin{table}[t]
\centering
\small
\caption{Task configurations for reproduction. The high-fidelity budget
is 300 epochs in every setting. Low gives the screening budget in epochs;
$\delta^\star$ is the promotion noise floor and Margin is the
low-fidelity refinement threshold.}
\label{tab:task_calibration}
\begin{tabular*}{\columnwidth}{@{\extracolsep{\fill}}llrrr@{}}
\toprule
Dataset & Method & Low & $\delta^\star$ & Margin\\
\midrule
\multirow{4}{*}{CDs\&Vinyl} & NCF & 200 & 0.0030 & 0.0050\\
& NGCF & 150 & 0.0015 & 0.0013\\
& SASRec & 150 & 0.0030 & 0.0050\\
& GRU4Rec & 200 & 0.0030 & 0.0050\\
\midrule
\multirow{4}{*}{ML-1M} & NCF & 250 & 0.0020 & 0.0050\\
& NGCF & 150 & 0.0021 & 0.0045\\
& SASRec & 295 & 0.0007 & 0.0024\\
& GRU4Rec & 150 & 0.0005 & 0.0007\\
\midrule
\multirow{4}{*}{Baby} & NCF & 15 & 0.0008 & 0.0026\\
& NGCF & 115 & 0.0005 & 0.0016\\
& SASRec & 25 & 0.0015 & 0.0050\\
& GRU4Rec & 25 & 0.0004 & 0.0012\\
\midrule
\multirow{4}{*}{Software} & NCF & 130 & 0.0011 & 0.0038\\
& NGCF & 140 & 0.0006 & 0.0019\\
& SASRec & 40 & 0.0006 & 0.0019\\
& GRU4Rec & 95 & 0.0004 & 0.0013\\
\bottomrule
\end{tabular*}
\end{table}

\paragraph{Task-specific calibration.}
Training budgets are epochs, with a high-fidelity cap of 300. Both the
promotion floor $\delta^\star$ and refinement margin in
Table~\ref{tab:task_calibration} are absolute differences in the validation
objective of Eq.~\eqref{eq:validation-objective}. The former enters the
cost-adjusted admission threshold; the latter controls the low-fidelity
progress reference and backtracking in Section~\ref{sec:method:screen}.
The analyst receives the same noise floor as the gate.

The automatic calibration protocol profiles the initial mechanism with
seeds 12352, 12353, and 12354 and three additional repetitions of seed 12352,
giving six runs and four observations at the repeated seed. It records
validation curves every five epochs and uses the best score attained
within each budget. For each curve, the peak epoch is the first epoch
attaining its maximum. The low rung is 1.35 times the latest such epoch,
rounded to the nearest evaluation interval and capped one interval below
the high rung. If a baseline peak is at the cap, the procedure uses half
the high budget. At the chosen rungs, it estimates sample standard
deviations $s_L$ and $s_H$ from fixed-seed repetitions. If either repeated-run
deviation is measured as zero, both are replaced by their corresponding
seed-to-seed deviations.
The policy sets $\delta^\star=Q(0.25\sqrt{2}\,s_H)$ and the refinement
margin to $Q(1.2s_L)$, where $Q$ rounds to the nearest $10^{-4}$ and is
bounded below by $10^{-4}$. The four CDs\&Vinyl settings and ML-1M/NCF
and ML-1M/NGCF use their earlier fixed configurations; the remaining
settings use the automatic protocol. Reproduction uses the listed
constants throughout search.

\paragraph{Process rubric.}
Section~\ref{sec:method:reward} defines the 5, 3, 6, and 4 criteria.
The trace supports each binary hit. The first three dimensions average
across experiments; novelty uses the complete round, preceding rounds,
and recorded failures in Eq.~\eqref{eq:novelty-score}.

\paragraph{Outside wiki.}
\label{app:outside-wiki}

For framework comparisons, OE, DE, and DualEvolve share the same prebuilt,
read-only knowledge snapshot. Recommendation papers are screened offline
using keywords and the relevance of their titles and abstracts.
Paper Fact Cards record source mechanisms and implementation
notes; cards covering a shared technique form Method Prior Cards with
source links. The archived export contains 111 method cards, with 108
indexed entries passing a seven-family keyword map for retrieval eligibility.

Within DualEvolve, at each refinement step, model status and the preceding diagnosis query
BM25 ($k_1=1.5$, $b=0.75$), retrieving up to 12 positively scored methods.
A frozen LLM selects up to five pool entries; unparseable or unavailable
selection falls back to BM25 top entries. The initial seed step does not
query the index. The top one or two selected methods receive additional
reading of linked Paper Fact Cards. The proposer combines their mechanisms
and parameter guidance with the diagnosis, and the coder adapts them through
the editable interface in Section~\ref{sec:problem:objects}.

A prebuilt snapshot is loaded before each search and remains read-only,
with source ingestion and maintenance disabled. Queries and candidate
implementations can change as diagnoses evolve. Task outcomes and
consolidated lessons enter the inside wiki; candidate admission uses the
fixed evaluator and promotion gate.

\paragraph{Mechanism changes.}
The archived final mechanisms include both model changes and training
changes (Table~\ref{tab:mechanisms}). Thirteen have a within-lineage
refinement as their final recorded origin and three have a newly selected
starting point. A final mechanism can combine this transition with
modifications accumulated in preceding rounds.

\section{Additional Experimental Evidence}
\label{app:additional-evidence}

\paragraph{Screened candidates at full budget.}
Across the 16 search traces, the candidate with the highest high-fidelity
score ranks first at low fidelity in 7 settings and between second and
tenth in the remainder. This ranking variation motivates the separation
of low-fidelity screening from high-fidelity selection.
Table~\ref{tab:screening_details} reports the 67 screened candidates
from three settings, re-evaluated with their original configurations,
seed 12345, and each
source search's full training budget. The comparison uses the validation objective in
Eq.~\eqref{eq:validation-objective} and each source search's recorded
high-fidelity winner. Three re-evaluation attempts terminate
unsuccessfully. Two candidates match their reference winner exactly.
One screened candidate on Software/\allowbreak GRU4Rec was initially below
the low-fidelity winner but rose from 0.4553 at low fidelity to 0.5987 at
full fidelity, exceeding the recorded winner by 0.0074. Among the 67
screened candidates, one therefore overtakes its recorded search winner
after longer training. These outcomes characterize the tradeoff between
full-budget training cost and candidate recall.

\begin{table}[H]
\centering
\caption{Full-budget outcomes of screened candidates. ``Overtake'' and
``Tie'' compare validation scores with the corresponding search winner.}
\label{tab:screening_details}
\begin{tabular*}{\columnwidth}{@{\extracolsep{\fill}}lrrr@{}}
\toprule
Setting & Count & Overtake & Tie \\
\midrule
CDs\&Vinyl/GRU4Rec & 19 & 0 & 2 \\
Baby/SASRec & 30 & 0 & 0 \\
Software/GRU4Rec & 18 & 1 & 0 \\
\midrule
Total & 67 & 1 & 2 \\
\bottomrule
\end{tabular*}
\end{table}

\paragraph{Search-level statistics for significance tests.}
Each search observation averages its selected mechanism's test scores over
seeds 12352--12354. OE, DE, and \method{} run independently under the same
three search-seed identifiers. Table~\ref{tab:search_repeats} gives all means
and \method{} SDs; Table~\ref{tab:comparator_search_sd} supplies OE/DE SDs.
Welch tests compare against the higher-mean OE/DE method.

\begin{table}[H]
\centering
\small
\caption{Comparator sample SDs over three independent searches.
Table~\ref{tab:search_repeats} reports all means and \method{} SDs.
Together, these statistics supply the Welch tests against the higher-mean
OE/DE comparator.}
\label{tab:comparator_search_sd}
\begin{tabular*}{\columnwidth}{@{\extracolsep{\fill}}lrrrr@{}}
\toprule
Method & H@5 & H@10 & N@5 & N@10 \\
\midrule
\multicolumn{5}{c}{Baby/NGCF} \\
\midrule
OE & 0.0054 & 0.0031 & 0.0034 & 0.0026 \\
DE & 0.0028 & 0.0052 & 0.0037 & 0.0038 \\
\midrule
\multicolumn{5}{c}{Software/NGCF} \\
\midrule
OE & 0.0046 & 0.0032 & 0.0029 & 0.0051 \\
DE & 0.0053 & 0.0044 & 0.0026 & 0.0045 \\
\midrule
\multicolumn{5}{c}{CDs\&Vinyl/GRU4Rec} \\
\midrule
OE & 0.0033 & 0.0030 & 0.0027 & 0.0036 \\
DE & 0.0040 & 0.0054 & 0.0045 & 0.0045 \\
\bottomrule
\end{tabular*}
\end{table}

\paragraph{Human annotations and cross-model grading.}
The original process grader is GLM-5.2.
Four annotators independently labeled the sampled traces using the process
rubric in Section~\ref{sec:method:reward}. Each human reference label was
determined by a majority of at least three of the four annotators.
The annotation sample comprises 33 decision-quality units,
34 experimental-rigor units, and 33 analysis-depth units, drawn
across the 16 settings and search rounds. Each unit covers the seven
candidate experiments from one round for one scoring dimension.
These dimensions contain
five, three, and six criteria, respectively. Their 3,255 individual
judgments are matched to the original grader's criterion hits and
supporting evidence. For agreement statistics, each original-grader
criterion label is the majority vote of its three grading passes.
Novelty is evaluated at the round level: 33 sampled
rounds are each annotated against its four criteria, adding 132 judgments.
The complete reference set therefore contains 3,387 criterion labels. Table~\ref{tab:grader_details}
reports agreement with these majority-vote human reference labels. Cohen's $\kappa$ measures
chance-corrected grader--human agreement, with the overall row pooling the
candidate-level and round-level judgments.

The alternative-grader evaluation uses identical prompts on the same
100 annotation units for decision quality, experimental rigor, and
analysis depth, with three grading passes per unit and model and a criterion
marked positive by majority vote. It comprises 600 completed calls.
DeepSeek-V4-Flash has 85.9--87.3\% agreement with the original
grader across the three grading passes, while Qwen3.5-Plus has
81.2--81.7\% agreement. The original grader marks 2,658 criterion
judgments positive, compared with 2,999 for Qwen's majority vote.
For human-reference comparisons, agreement is the fraction of matching criterion labels;
Cohen's $\kappa$ accounts for chance agreement under the label distributions. DeepSeek-V4-Flash and Qwen3.5-Plus reach similar human
agreement rates (81.5\% and 81.6\%), with $\kappa$ values of 0.378 and 0.202.
Table~\ref{tab:grader_details} reports the matching proportions and
chance-corrected agreement for each dimension and grader.

\begin{table}[H]
\centering
\caption{Grader agreement with human annotations. Upper rows report the
original grader across candidate- and round-level criteria; lower rows report
alternative graders' majority votes on the three candidate-level dimensions.
Cohen's $\kappa$ corrects for chance agreement.}
\label{tab:grader_details}
\begin{tabular*}{\columnwidth}{@{\extracolsep{\fill}}lccc@{}}
\toprule
Dimension or grader & Agreement & Cohen's $\kappa$ & Level \\
\midrule
Decision quality & 80.8\% & 0.271 & Candidate \\
Experimental rigor & 76.6\% & 0.333 & Candidate \\
Analysis depth & 87.7\% & 0.598 & Candidate \\
Novelty & 89.4\% & 0.667 & Round \\
All annotated dimensions & 83.1\% & 0.432 & Mixed \\
\midrule
DeepSeek-V4-Flash & 81.5\% & 0.378 & Candidate \\
Qwen3.5-Plus & 81.6\% & 0.202 & Candidate \\
\bottomrule
\end{tabular*}
\end{table}

\paragraph{Additional searches with fixed rules.}
The four searches in Table~\ref{tab:fixed_rules_additional} fix
all harness and skill updates, including the redirection guidance normally
written to the harness after one non-promoting round. Inner evolution,
task-specific memory, process grading, and posterior updates remain active,
and the same two-round early-stopping condition remains in effect. The
searches use the frozen GLM-5.2 and a four-round cap. The table reports each
selected mechanism's search
score and its validation score after retraining with seed 12352 at
full fidelity. Three searches select revised
mechanisms; Software/NCF terminates after two rounds while retaining
the initial mechanism. The results describe the behavior of inner
evolution under a fixed control layer across collaborative filtering
and sequential recommendation.

\begin{table}[H]
\centering
\caption{Additional Fixed Rules searches. Search and re-evaluation scores
use the validation objective. Re-evaluation uses seed 12352 and the full
training budget; rounds denote completed search rounds.}
\label{tab:fixed_rules_additional}
\begin{tabular*}{\columnwidth}{@{\extracolsep{\fill}}lrrr@{}}
\toprule
Setting & Rounds & Search & Re-eval. \\
\midrule
Baby/SASRec & 4 & 0.2621 & 0.2564 \\
Software/NCF & 2 & 0.4829 & 0.4869 \\
ML-1M/GRU4Rec & 4 & 0.6895 & 0.6945 \\
ML-1M/SASRec & 4 & 0.6967 & 0.6949 \\
\bottomrule
\end{tabular*}
\end{table}

\paragraph{Clean evaluation after shortcut exposure.}
After removing the shortcut interfaces, we re-evaluate each selected
mechanism with seeds 12352, 12353, and 12354. Table~\ref{tab:stress_clean}
reports the search score and clean validation and test means. The first
fully disclosed run depends on the removed interfaces and cannot execute;
the other five obtain clean validation scores between 0.2680 and 0.2769.

\begin{table}[H]
\centering
\small
\caption{Clean evaluation of shortcut-exposure winners. U, P, and F denote
undisclosed, partial, and full disclosure. Reported and Clean are validation
objectives; H@5 and N@5 are three-seed test means. A dash denotes failure
without the shortcut interfaces.}
\label{tab:stress_clean}
\begin{tabular*}{\columnwidth}{@{\extracolsep{\fill}}lrrrr@{}}
\toprule
Run & Reported & Clean & H@5 & N@5 \\
\midrule
U/1 & 0.2625 & 0.2769 & 0.3025 & 0.2060 \\
P/1 & 0.2668 & 0.2697 & 0.2948 & 0.2017 \\
F/1 & 1.6961 & --- & --- & --- \\
U/2 & 0.2436 & 0.2732 & 0.2987 & 0.2040 \\
P/2 & 0.2743 & 0.2680 & 0.2937 & 0.1999 \\
F/2 & 0.2635 & 0.2684 & 0.2982 & 0.2043 \\
\bottomrule
\end{tabular*}
\end{table}

\section{Transferred Rules and Search Cost}
\label{app:transfer_cost}

The target-specific Full run starts from the initial orchestration and
explores simplified graph propagation, feature transformations, and
aggregation. The transferred Baby/NCF rules guide a different candidate
sequence in which a stronger mechanism emerges through target evaluation.
The matched No Transfer control uses the initial rules under the same
one-round target-side schedule. Table~\ref{tab:transfer_knowledge} connects
representative source-evolved guidance to target revisions.

DE uses a research, planning, and report-generation workflow for candidate
production. \method{} concentrates later calls on diagnosis and revision
using accumulated experimental evidence.

\end{document}